\documentclass[12pt, a4paper]{article}
\usepackage{graphicx}
\usepackage{amssymb}
\usepackage{amsmath}
\usepackage{bm}
\usepackage{color}
\usepackage{theorem}
\usepackage{subcaption}
\usepackage{colortbl}
\usepackage{listings}
\usepackage{tabularx}
\usepackage{longtable}
\usepackage[utf8]{inputenc}

\usepackage[sort&compress,numbers, merge]{natbib}

\definecolor{Orange}{cmyk}{0,0.61,0.87,0}
\definecolor{JungleGreen}{cmyk}{0.99,0,0.52,0}
\definecolor{OliveGreen}{cmyk}{0.64,0,0.95,0.40}
\definecolor{Brown}{cmyk}{0,0.70,1,0.40}
\definecolor{RoyalBlue}{cmyk}{0.71,0.53,0,0.12}
\definecolor{Gray}{cmyk}{0,0,0,0.40}
\definecolor{LightPink}{cmyk}{0.0,0.25,0,0}
\definecolor{LLightPink}{cmyk}{0.0,0.10,0,0}
\definecolor{LightBlue}{cmyk}{0.25,0,0,0}
\definecolor{LightGray}{cmyk}{0,0,0,0.2}

\newcommand{\Slash}[1]{{\ooalign{\hfil/\hfil\crcr$#1$}}}

\allowdisplaybreaks[1]

\usepackage[colorlinks=true, linkcolor=OliveGreen, citecolor=RoyalBlue,
urlcolor=RoyalBlue]{hyperref}

\begin{document}

\begin{titlepage}

\begin{flushright}
{\tt 
KIAS-P19012, 
UT-19-02 \\
FTPI-MINN-19/08,
UMN-TH-3817/19
}
\end{flushright}

\vskip 1.35cm
\begin{center}

{\large
{\bf
A Minimal SU(5) SuperGUT in Pure Gravity Mediation
}
}

\vskip 1.5cm

Jason L. Evans$^{a}$,
Natsumi Nagata$^{b}$, 
and
Keith A. Olive$^{c}$

\vskip 0.8cm

{\it $^a$School of Physics, KIAS, Seoul 130-722, Korea} \\[3pt]
{\it $^b$Department of Physics, University of Tokyo, Bunkyo-ku, Tokyo
 113--0033, Japan} \\[3pt]
{\it $^c$William I. Fine Theoretical Physics Institute, School of
 Physics and Astronomy, \\ University of Minnesota, Minneapolis,
 Minnesota 55455, USA} 

\date{\today}

\vskip 1.5cm

\begin{abstract}
The lack of evidence for low-scale supersymmetry suggests that the scale of supersymmetry breaking may be higher than originally anticipated. However, there remain many motivations for supersymmetry including gauge coupling unification and a stable dark matter candidate. Models like pure gravity mediation (PGM) evade LHC searches while still providing a good dark matter candidate and gauge coupling unification. Here, we study the effects of PGM if the input boundary conditions for soft supersymmetry breaking masses are pushed beyond the unification scale and higher dimensional operators are included.  The added running beyond the unification scale opens up the parameter space by relaxing the constraints on $\tan\beta$. If higher dimensional operators involving the SU(5) adjoint Higgs are included, the mass of the heavy gauge bosons of SU(5) can be suppressed leading to proton decay, $p\to \pi^0 e^+$, that is within reach of future experiments. Higher dimensional operators involving the supersymmetry breaking field can generate additional contributions to the A- and B-terms of order $m_{3/2}$. The threshold effects involving these A- and B-terms significantly impact the masses of the gauginos and can lead to a bino LSP. In some regions of parameter space the bino can be degenerate with the wino or gluino and give an acceptable dark matter relic density.

\end{abstract}

\end{center}
\end{titlepage}

\section{Introduction}

Despite its many motivations, low energy supersymmetry (SUSY) ($E < 1$ TeV)
is yet to be discovered at the LHC \cite{nosusy} calling into question the scale of supersymmetry breaking. 
While it is possible that the discovery of supersymmetry is just around the corner for the LHC, it is also 
quite possible that supersymmetry breaking lies at higher or even much higher energy scales.  For example,
in models of split supersymmetry \cite{split}, 
scalar masses may lie beyond the PeV scale.
In minimal anomaly mediated supersymmetry breaking (mAMSB)
\cite{anom,ggw,mAMSB}, the gravitino
is also of order a PeV, while 
 scalar masses are set independently and may lie somewhere in between the TeV and PeV mass scales \cite{mc13,ehow++}. Since gaugino masses are
generated at the 1-loop level, their masses are of order
1 TeV.  In pure gravity mediation (PGM) \cite{pgm,eioy,eioy2, Evans:2014xpa,eioy5}, a variant of split supersymmetry, scalar masses are
set by the gravitino mass as in models of minimal supergravity \cite{bfs} and are near the PeV scale,
while gaugino masses are generated by anomaly mediation.
In both mAMSB and PGM, the lightest supersymmetric particle (LSP) is often the wino\footnote{It is also 
possible that the LSP is a Higgsino in PGM models \cite{eioy5}.}. The scale of supersymmetry breaking can be even higher as in high-scale supersymmetry 
\cite{hssusy,hssusy2,egko}. It is in fact possible
that the scale of supersymmetry breaking lies
beyond the inflationary scale leaving behind only the 
gravitino with a mass of order 1 EeV \cite{eev}. In the most extreme case, supersymmetry breaking occurs at the string or Planck scale and does 
not play a role in low energy phenomenology.

Here, we take a more optimistic view. 
While some of the supersymmetric spectrum may be heavy,
part of it may remain light and accessible to experiment.
In conventional models of supersymmetry based on 
supergravity such as the constrained minimal supersymmetric standard model (CMSSM) \cite{cmssm,ELOS,eelnos,azar,eeloz,ehow++}, the soft masses lie below about 10 TeV. In these models, some form of tuning of its input parameters is required
to obtain the needed mass degeneracies which allow the relic density to fall into the range determined by
CMB experiments \cite{planck18}. 
In PGM models, the gauginos remain light. The dark matter candidate is the wino and the mass degeneracies that set the relic density are enforced by the SU(2) gauge symmetry. 

In all of the models we have been referring to,
there are strong relations among the supersymmetry breaking parameters at some very high energy scale.
In the CMSSM, for example, there is a common universality scale (taken to be the grand unified theory (GUT) scale, $M_{\rm GUT}$, often 
defined as the renormalization scale where the two 
electroweak gauge couplings are equal). At this scale,
all soft scalar masses are equal to a common mass
$m_0$, the three gaugino masses are set to $m_{1/2}$,
supersymmetry breaking tri-linear terms are all set to 
$A_0$. The $\mu$-term and the supersymmetry breaking 
bilinear, $B$ are set at the weak scale by the minimization of the Higgs potential, leaving the ratio
of the two Higgs expectation values, $\tan \beta$ as a 
fourth free parameter. However, there is no firm
reason that the universality scale for supersymmetry breaking must lie at the GUT scale. That scale may lie 
below the GUT scale as in subGUT models \cite{subGUT,ELOS,eelnos,eeloz,MCsubGUT} or above the 
GUT scale as in the superGUT models discussed here \cite{superGUT,emo,dlmmo,azar,eenno,Evans:2018ewb}. Both 
subGUT and superGUT models necessarily introduce at 
least one additional parameter (over the CMSSM),
namely the universality input scale, $M_{\rm in}$.
While superGUT models also introduce new parameters
directly associated with the specific GUT, the 
additional running between $M_{\rm in}$ and $M_{\rm GUT}$
offers additional flexibility due to the 
non-universality of the supersymmetry breaking parameters at the GUT scale. 

While the CMSSM is highly constrained with only 4 parameters, minimal supergravity (mSUGRA), mAMSB and PGM have even fewer free parameters. In mSUGRA models, 
the $B$-term is fixed at the GUT scale, by $B_0 = A_0 - m_0$, and so $\tan \beta$ must be determined by the Higgs minimization conditions together with the $\mu$-term \cite{vcmssm}. 
mAMSB models also have three free parameters which are often chosen to be $m_0$, $\tan \beta$ and the gravitino mass, $m_{3/2}$, with the the gaugino masses and $A$-terms determined from $m_{3/2}$. In principle,
PGM models have only one free parameter, $m_{3/2}$.
As in the case of mAMSB, gaugino masses and $A$-terms are determined from $m_{3/2}$, $m_0 = m_{3/2}$, $B_0 = A_0 - m_0 \approx -m_0$, and as in mSUGRA models, $\tan \beta$ and $\mu$ are fixed at the weak scale \cite{eioy}. However,
in most cases, this one-parameter model is too restrictive and can be relaxed by adding a Giudice-Masiero term \cite{gm,Inoue:1991rk, Casas:1992mk, dlmmo} which allows one to choose $\tan \beta$ as a second free parameter. 

As is well known, one of the prime motivations for 
supersymmetry is grand unification. The problem of 
gauge coupling unification \cite{Dimopoulos:1981yj} and the gauge hierarchy \cite{Maiani:1979cx}
problems are both relaxed in supersymmetric models. 
However in the context of minimal supersymmetric 
SU(5) \cite{Dimopoulos:1981zb}, a new problem arises
due to proton decay from dimension-5 operators \cite{dim5protondecay}. In \cite{Murayama:2001ur},
it was argued that low energy supersymmetric models
were incompatible with minimal SU(5). These arguments
have been relaxed, however, as the scale of 
supersymmetry breaking is pushed past the TeV
scale \cite{Hisano:2013exa, McKeen:2013dma, Nagata:2013sba, Nagata:2013ive,
Evans:2015bxa, eelnos}. They are further relaxed in
PGM and other models with high scale supersymmetry breaking \cite{Evans:2015bxa}.

Previously, we considered PGM models in the context
of minimal SU(5) with supersymmetry breaking universality input at the GUT scale \cite{Evans:2015bxa}. To relax the constraint on 
$\tan \beta$ (which must be around 2 to achieve
radiative electroweak symmetry breaking (rEWSB)), we
introduced some non-universality in the Higgs soft masses. This allowed for higher $\tan \beta$, a heavier
Higgs mass (with better compatibility with experiment),
and a proton lifetime within reach of on-going and future experiments. In the models considered,
the gravitino mass was of order 100 TeV and as a result, the wino mass was too small to sustain the 
needed relic density of dark matter. Here, we
consider a superGUT version of PGM models. The additional running between $M_{\rm in}$ and $M_{\rm GUT}$, which as noted earlier, generates some non-universalities among the soft masses at the GUT scale. In addition,
we consider the effects of higher dimension operators which may further affect the gaugino mass spectrum.
The effects of these operators may change the identity of the lightest gaugino from wino to bino and even a gluino. We find that both wino-bino co-annihilations \cite{Baer:2005jq, Harigaya:2014dwa, Nagata:2015pra} and bino-gluino co-annihilations \cite{Profumo:2004wk, deSimone:2014pda, Harigaya:2014dwa, Evans:2014xpa,
Ellis:2015vaa, Nagata:2015hha} may play an important role in determined the cold dark matter density in these models.  These higher dimensional operators affect the masses of the SU(5) particles. In fact, these operators also allow for a lighter SU(5) gauge boson putting dimension-6 proton decay within reach of future experiments. 

The paper is organized as follows.
In the next section, we describe the minimal SU(5) model, the boundary conditions at the input scale, $M_{\rm in}$ set by PGM, the higher dimension operators we consider, and the matching conditions at the GUT scale between SU(5) parameters and those associated with the Standard Model (SM).  In section \ref{sec:protondecay}, we discuss our
calculation of the proton lifetime. In section \ref{sec:result}, we
present results for superGUT PGM models. Our conclusions are given in section \ref{sec:conclusion}.

\section{Model}
\label{sec:model}

We restrict our attention to a minimal SU(5) superGUT model in the context of PGM \cite{pgm,eioy,eioy2, Evans:2014xpa,eioy5}. Above the GUT scale, the field content is that of  minimal SUSY SU(5)~\cite{Dimopoulos:1981zb}, which is briefly reviewed in
Sec.~\ref{sec:minimalsu5}. In our model, the soft SUSY-breaking
parameters are generated by PGM
 at a scale $M_{\rm in} \ge M_{\rm GUT}$, as in the superGUT models discussed in
Refs.~\cite{superGUT,emo,dlmmo, azar, eenno,
Evans:2018ewb}. We discuss this framework in Sec.~\ref{sec:spgm}.  
Generically speaking, we expect the theory to contain higher-dimensional
effective operators suppressed by the Planck scale. We consider the
possible effects of such operators in Sec.~\ref{sec:hdop}. We then
show the GUT-scale matching conditions in the presence of these
effective operators in Sec.~\ref{sec:matching}. Finally, in
Sec.~\ref{sec:modelsummary}, we summarize the setup we analyze in the
following sections.

\subsection{Minimal SUSY SU(5) GUT}
\label{sec:minimalsu5}

The minimal SUSY SU(5) GUT \cite{Dimopoulos:1981zb} consists of three
$\bf{\overline{5}}$ (${\Phi}_i$) and $\bf{10}$ (${\Psi}_i$) matter
chiral superfields with $i = 1,2,3$ the generation index, an adjoint
GUT Higgs superfield $\Sigma \equiv \sqrt{2}\Sigma^A T^A$ with $T^A$
($A=1, \dots, 24$) the generators of SU(5), 
and a pair of the Higgs chiral superfields in  ${\bf 5}$ and $\overline{\bf 5}$
representations, $H$ and
$\overline{H}$, respectively. The three generations of the MSSM matter
fields are embedded into $\Phi_i$ and $\Psi_i$ as in the original
Georgi-Glashow model \cite{Georgi:1974sy}, while the MSSM Higgs chiral
superfields $H_u$ and $H_d$ are in $H$ and $\overline{H}$ accompanied
by the ${\bf 3}$ and $\overline{\bf 3}$ color Higgs superfields $H_C$
and $\overline{H}_C$, respectively. The renormalizable superpotential in
this model is then given by
\begin{align}
 W_5 &=  \mu_\Sigma {\rm Tr}\Sigma^2 + \frac{1}{6} \lambda^\prime {\rm
 Tr} \Sigma^3 + \mu_H \overline{H} H + \lambda \overline{H} \Sigma H
\nonumber \\
&+ \left(h_{\bf 10}\right)_{ij} \epsilon_{\alpha\beta\gamma\delta\zeta}
 \Psi_i^{\alpha\beta} \Psi^{\gamma\delta}_j H^\zeta +
 \left(h_{\overline{\bf 5}}\right)_{ij} \Psi_i^{\alpha\beta} \Phi_{j \alpha}
 \overline{H}_\beta ~,
\label{W5}
\end{align}
where the Greek sub- and super-scripts denote SU(5) indices, and
$\epsilon_{\alpha \beta \gamma \delta \zeta}$ is the totally
antisymmetric tensor with $\epsilon_{12345}=1$. $R$-parity is
assumed to be conserved in this model.

The adjoint Higgs $\Sigma$ is assumed to have a vacuum expectation value
(VEV) of the form
\begin{equation}
 \langle \Sigma \rangle = V \cdot {\rm diag} \left(2,2,2,-3,-3\right) ~,
\end{equation}
with 
\begin{equation}
 V = \frac{4\mu_\Sigma}{\lambda^\prime}~.
\end{equation}
This VEV breaks the SU(5) GUT group into the SM gauge group,
$\text{SU}(3)_C \otimes \text{SU(2)}_L \otimes \text{U}(1)_Y$, while
giving masses to the GUT gauge bosons
\begin{equation}
 M_X = 5 g_5 V~,
\label{eq:mx}
\end{equation}
with $g_5$ the
SU(5) gauge coupling constant. In addition, we impose the fine-tuning
condition $\mu_H -3\lambda V \lesssim {\cal O}(m_{3/2})$ to ensure the
doublet-triplet mass splitting in $H$ and $\overline{H}$. The color and
weak adjoint components of $\Sigma$, the singlet component of $\Sigma$,
and the color-triplet Higgs states then acquire masses
\begin{align}
 M_\Sigma = \frac{5}{2} \lambda^\prime V ~, 
\qquad
 M_{\Sigma_{24}} = \frac{1}{2} \lambda^\prime V ~, 
\qquad
 M_{H_C} = 5\lambda V ~,
\label{eq:msigma}
\end{align}
respectively.

\subsection{SuperGUT Pure Gravity Mediation}
\label{sec:spgm}

In the minimal SUSY SU(5) GUT, the soft SUSY-breaking terms are given by 
\begin{align}
 {\cal L}_{\rm soft} = &- \left(m_{\bf 10}^2\right)_{ij}
 \widetilde{\psi}_i^* \widetilde{\psi}_j
- \left(m_{\overline{\bf 5}}^2\right)_{ij} \widetilde{\phi}^*_i
 \widetilde{\phi}_j
- m_H^2 |H|^2 -m_{\overline{H}}^2 |\overline{H}|^2 - m_\Sigma^2 {\rm Tr}
\left(\Sigma^\dagger \Sigma\right)
\nonumber \\
&-\biggl[
\frac{1}{2}M_5 \widetilde{\lambda}^{A} \widetilde{\lambda}^A
+ A_{\bf 10} \left(h_{\bf 10}\right)_{ij}
 \epsilon_{\alpha\beta\gamma\delta\zeta} \widetilde{\psi}_i^{\alpha\beta}
 \widetilde{\psi}^{\gamma\delta}_j H^\zeta
+ A_{\overline{\bf 5}}\left(h_{\overline{\bf 5}}\right)_{ij}
 \widetilde{\psi}_i^{\alpha\beta} \widetilde{\phi}_{j \alpha}  \overline{H}_\beta
\nonumber \\
&+ B_\Sigma \mu_\Sigma {\rm Tr} \Sigma^2 +\frac{1}{6} A_{\lambda^\prime
 } \lambda^\prime  {\rm Tr} \Sigma^3 +B_H \mu_H \overline{H} H+
 A_\lambda \lambda \overline{H} \Sigma H +{\rm h.c.}
 \biggr]~,
\end{align}
where $\widetilde{\psi}_i$ and $\widetilde{\phi}_i$ are the scalar
components of $\Psi_i$ and $\Phi_i$, respectively,
the $\widetilde{\lambda}^A$ are the SU(5) gauginos, and for the scalar
components of the Higgs superfields we use the same symbols as for the
corresponding superfields. In this work, we assume that these soft
SUSY-breaking terms in the visible sector are induced at a scale $M_{\rm
in} > M_{\rm GUT}$ through PGM
\cite{pgm,eioy,eioy2}. We focus on the minimal PGM
content for the moment, and discuss the case with the Planck-scale
suppressed non-renormalizable operators in the subsequent subsection.

In minimal PGM, the K\"{a}hler potential is
assumed to be flat, and hence the soft scalar masses are universal and
equal to the gravitino mass $m_{3/2}$, as in mSUGRA
\cite{bfs,vcmssm}: 
\begin{align}
  \left(m_{\bf 10}^2\right)_{ij} =
\left(m_{\overline{\bf 5}}^2\right)_{ij}
\equiv m_{3/2}^2 \, \delta_{ij} ~,
\qquad
m_H = m_{\overline{H}} = m_\Sigma \equiv m_{3/2} ~.
\end{align}
On the other hand, the gaugino masses and $A$-terms vanish at tree
level---such a situation is naturally obtained if there is no singlet
SUSY breaking field and/or in models with strong moduli stabilization
\cite{Dudas:2006gr, Abe:2006xp, Kallosh:2011qk, Linde:2011ja, dlmmo}. In
this case they are induced by anomaly mediation
\cite{anom} at the quantum level, and are thus
suppressed by a loop factor. The contribution of the one-loop suppressed
$A$-terms to the physical observables is insignificant and thus we can
safely neglect them in the following discussion: 
\begin{equation}
 A_{\bf 10} \simeq A_{\overline{\bf 5}} \simeq A_{\lambda} \simeq
  A_{\lambda^\prime} \simeq 0 ~.
\label{eq:aanom}
\end{equation}
For gaugino masses, on
the other hand, AMSB generates
\begin{equation}
 M_5 = \frac{b_5 g_5^2}{16\pi^2} m_{3/2} ~,
\label{eq:m5anom}
\end{equation}
with $b_5 = -3$ the beta-function coefficient of the SU(5) gauge
coupling. Thus, the gauginos have a universal mass above the GUT scale,
which is orders-of-magnitude smaller than the scalar masses. 

In addition to these contributions, in PGM,
Giudice-Masiero (GM) terms \cite{gm, Inoue:1991rk, Casas:1992mk, dlmmo, eioy, eioy2, eenno}
are added to the K\"{a}hler potential: 
\begin{equation}
 \Delta K =  c_\Sigma {\rm Tr}(\Sigma^2) + c_H H \overline{H}+
  \text{h.c.} 
\label{eq:gm}
\end{equation} 
These terms shift the corresponding $\mu$-terms as
\begin{align}
 \mu_\Sigma &\to \mu_\Sigma + c_\Sigma\, m_{3/2}~, \nonumber \\
 \mu_H &\to \mu_H + c_H \,m_{3/2}~,
\label{eq:gmmu}
\end{align}
and generate $B$-terms
\begin{align}
 B_\Sigma \,\mu_\Sigma &= - m_{3/2} \, \mu_\Sigma 
+ 2 c_\Sigma \, m_{3/2}^2 ~, \nonumber \\
 B_H \,\mu_H &= - m_{3/2} \, \mu_H
+ 2 c_H \, m_{3/2}^2 ~, 
\label{eq:banom}
\end{align}
where the first terms correspond to the usual supergravity
contribution. We note that the second terms on the right-hand side of
the above equations are smaller than the first terms by ${\cal
O}(m_{3/2}/M_{\rm GUT})$. For the contribution to the $\mu$-terms
\eqref{eq:gmmu}, therefore, we can safely neglect this in the following
discussion. As for the $B$-terms \eqref{eq:banom}, on the other hand,
these terms do play a role in assuring successful electroweak symmetry breaking,
as we will see in Sec.~\ref{sec:matching}.

\subsection{Planck-scale suppressed higher-dimensional operators}
\label{sec:hdop}

GUT phenomenology has some sensitivity to the Planck-scale
suppressed higher-dimensional operators since the GUT scale is only
about two orders of magnitude lower than the Planck scale.
The SUSY spectrum may also be affected by non-renormalizable operators that
consist of both visible and SUSY-breaking sector fields. In this subsection, we
discuss the effect of such operators.

\subsubsection{Non-renormalizable operators without SUSY-breaking
   fields}
\label{sec:nrowosbf}

We first discuss the effect of the Planck-scale suppressed
higher-dimensional operators which do not include SUSY-breaking
fields. Here, we mainly consider the dimension-five operators 
involving the
adjoint Higgs field $\Sigma$, as the effect of such
operators is suppressed only by a factor of ${\cal O}(V/M_P)$.

Among such operators, 
\begin{equation}
 W_{\rm eff}^{\Delta g} = \frac{c}{M_P} {\rm Tr}\left[
\Sigma {\cal W} {\cal W}
\right] ~,
\label{eq:SigmaWW}
\end{equation}
has the most significant effect on our analysis, where ${\cal W}\equiv
T^A {\cal W}^A$ denotes the superfields corresponding to the field
strengths of the SU(5) gauge vector bosons ${\cal V} \equiv {\cal V}^A
T^A$ and $M_P$ is the reduced Planck mass. This effective operator affects both
gauge coupling \cite{Ellis:1985jn, Hill:1983xh,
Tobe:2003yj} and gaugino mass \cite{Ellis:1985jn, Tobe:2003yj,
Anderson:1996bg} unification, as we see in detail in
Section~\ref{sec:matching}.

Another class of dimension-five operators that affect gauge coupling
unification is comprised of quartic superpotential terms of the
adjoint Higgs fields \cite{Bajc:2002pg}:
\begin{equation}
 W_{\rm eff}^\Sigma = \frac{c_{4\Sigma, 1}}{M_P} \left({\rm Tr} \Sigma^2\right)^2
+\frac{c_{4\Sigma, 2}}{M_P} {\rm Tr} \Sigma^4 ~.
\label{eq:sigma4}
\end{equation}
These operators can split the masses of the SU(3)$_C$ and SU(2)$_L$ adjoint
components in $\Sigma$, $M_{\Sigma_8}$ and $M_{\Sigma_3}$, by ${\cal
O}(V^2/M_P)$. This mass difference induces threshold corrections to gauge
coupling constants of  $\sim
\ln(M_{\Sigma_3}/M_{\Sigma_8})/(16\pi^2)$ \footnote{In general, this correction can lead to an enhanced color Higgs mass and so lengthens the dimension-5 mediated proton lifetime.  Since we are considering PGM, dimension-5 mediated proton decay is drastically suppressed making this correction less important.}. This effect can be
significant if $|\lambda^\prime| \lesssim |c_{4\Sigma, 1(2)}|V/M_P$. In
the following analysis, however, for simplicity, we assume that the contribution of
these operators is negligibly small.

The operators 
\begin{equation}
 W_{\rm eff}^{H \Sigma} = \frac{c_{H\Sigma, 1}}{M_P}\, \overline{H} \Sigma^2
  H  + \frac{c_{H\Sigma, 2}}{M_P}\, \overline{H} 
  H \,{\rm Tr} (\Sigma^2)~,
\label{eq:sigma2h2}
\end{equation}
are also of dimension-five. These operators shift the masses of the
SU(3)$_C$ triplet and SU(2)$_L$ doublet components of $H$ and
$\overline{H}$ by $\sim V^2/M_P$. In order to produce a MSSM Higgsino
mass term of order the SUSY-breaking scale, we need to cancel this
contribution by fine-tuning the parameter $\mu_H$.
Note that this does not introduce an additional fine tuning as, $\mu_H$ must be tuned in any case to cancel
$3\lambda V$ as noted earlier. Other than this, these operators provide 
no significant effect on our analysis.

The adjoint Higgs field may directly couple to the matter fields via
non-renormalizable operators such as
\begin{align}
  W_{\rm eff}^{\Delta h} &= \frac{c_{\Delta h,1 }^{ij}}{M_P} \Phi_{i \alpha}
  \Sigma^\alpha_{~\beta} \Psi^{\beta\gamma}_j \overline{H}_\gamma 
+\frac{c_{\Delta h, 2}^{ij}}{M_P} \Psi_i^{\alpha\beta} \Phi_{j\alpha}
\Sigma^\gamma_{~\beta} \overline{H}_\gamma \nonumber \\
&
+\frac{c_{\Delta h, 3}^{ij}}{M_P}
 \epsilon_{\alpha\beta\gamma\delta\zeta}
 \Psi_i^{\alpha\beta} \Psi^{\gamma\xi}_j \Sigma^\delta_{~\xi} H^\zeta 
+ \frac{c_{\Delta h, 4}^{ij}}{M_P}
\epsilon_{\alpha\beta\gamma\delta\zeta}
 \Psi_i^{\alpha\beta} \Psi^{\gamma\delta}_j \Sigma^\zeta_{~\xi} H^\xi  
~.
\label{eq:weffdelh}
\end{align}
After $\Sigma$ develops a VEV, these operators lead to Yukawa
interactions. Among these operators, the first one has a significant
implication for the prediction of the model, since it modifies the
unification of the down-type quark and charged-lepton Yukawa couplings,
which should be exact at the GUT scale in minimal SU(5). 
On the other hand, it is known that in most of the parameter space in
the MSSM, Yukawa unification is imperfect. For the third
generation, the deviation is typically at the ${\cal O}(10)$\% level,
while for the first two generations, there are ${\cal
O}(1)$ differences. These less successful predictions can be explained
if the first term in Eq.~\eqref{eq:weffdelh} is included \cite{nro,
Bajc:2002pg}. The change in the Yukawa couplings due to this operator is
${\cal O}(V/M_P) \sim 10^{-2}$. Therefore, if the value of the third
generation Yukawa coupling at the GUT scale is $\sim 10^{-1}$, this
contribution gives rise to an ${\cal O}(10)$\% correction. For the first
and second generation Yukawa couplings, this correction can be as large
as the values of the Yukawa couplings themselves since their values are
(much) smaller than $10^{-2}$. As we do not discuss Yukawa coupling
unification in this paper, we do not consider the operators in
Eq.~\eqref{eq:weffdelh} in what follows. 

We may also have K\"{a}hler-type operators of the form
\begin{equation}
 \Delta K_{\rm eff}^{F\Sigma} = \frac{c_{F\Sigma}}{M_P}F^* \Sigma F ~,
\end{equation}
where $F$ collectively stands for the fields in this model, $F = \Psi,
\Phi, H, \overline{H}, \Sigma$. These terms modify the wave functions of
the components in the field $F$ below the GUT scale. This correction is
insignificant for our analysis, and thus we neglect them in the
following discussions. 

There are lepton-number violating operators of the form
\begin{equation}
    W_{\rm eff}^{\scriptsize\Slash{L}}
    = \frac{c^{ij}_{\scriptsize\Slash{L}}}{M_P} \Phi_{i\alpha} \Phi_{j\beta} H^\alpha H^\beta ~.
    \label{eq:dim5lepv}
\end{equation}
These dimension-five operators give rise to Majorana masses for left-handed neutrinos at low energies, whose size is at most ${\cal O} (10^{-5})$~eV. This is much smaller than $\sqrt{m_{31}^2}$ and  $\sqrt{m_{21}^2}$ ($m_{31}^2$ and $m_{21}^2$ are the squared-mass differences of active neutrinos) observed in neutrino oscillation experiments, and therefore we need an extra source for neutrino mass generation. This can be provided by right-handed neutrinos, which are singlets under SU(5), as was considered for the superGUT CMSSM in \cite{Evans:2018ewb}. The presence of these additional singlet fields, as well as the operators in Eq.~\eqref{eq:dim5lepv}, does not affect the discussion below.

Finally, we may write down the dimension-five baryon- and lepton-number
violating operator,
\begin{equation}
 W_{\rm eff}^{\scriptsize\Slash{B}\Slash{L}}
= \frac{c^{ijkl}_{\scriptsize\Slash{B}\Slash{L}}}{M_P}
\epsilon_{\alpha\beta\gamma\delta\zeta} \Psi_i^{\alpha\beta}
\Psi_j^{\gamma\delta} \Psi_k^{\zeta \xi} \Phi_{l \xi} ~.
\label{eq:dim5prdecop}
\end{equation}
This operator yields the interactions of the form $QQQL$ and $\bar{U}
\bar{D} \bar{U} \bar{E}$, which induce proton decay such as $p \to K^+
\bar{\nu}$. If the coefficient of the operator \eqref{eq:dim5prdecop} is
${\cal O}(1)$, the presence of such an operator could be disastrous; as
shown in Ref.~\cite{Dine:2013nga}, in this case the proton lifetime is
predicted to be much shorter than the experimental bound even when the
SUSY-breaking scale is as high as $\sim 1$ PeV. This conclusion, however,
strongly depends on the flavor structure of the coefficient
$c^{ijkl}_{\scriptsize\Slash{B}\Slash{L}}$. For example, we may expect
there is a flavor symmetry at high energies that accounts for small
Yukawa couplings, as in the Froggatt-Nielsen model
\cite{Froggatt:1978nt}. In this case, the couplings
$c^{ijkl}_{\scriptsize\Slash{B}\Slash{L}}$ are strongly suppressed for
the first two generations, and the resultant proton decay rate can be
small enough to evade the experimental bound. In this work, we assume
that the coefficients of the operator \eqref{eq:dim5prdecop} are
sufficiently suppressed so that the proton decay rate induced by this
operator is negligibly small.

\subsubsection{Non-renormalizable operators with SUSY-breaking fields}
\label{sec:nrowsbf}

Next, we discuss non-renormalizable interactions that directly couple
the visible-sector fields to hidden-sector fields. The presence of
such interactions can modify the SUSY spectrum predicted in PGM. As we have seen in Sec.~\ref{sec:spgm}, in
PGM, gaugino masses are induced by quantum
effects and thus suppressed by a loop factor compared with the gravitino
mass $m_{3/2}$, which is the order parameter of SUSY
breaking. Thus, gaugino masses are more sensitive to the effects of
higher dimensional operators than scalar masses. Indeed, in Sec.~\ref{sec:matching} we will see
that if such higher-dimensional operators
generate $A$- and/or $B$-terms in the Higgs sector, they can
significantly modify the gaugino mass spectrum through threshold
corrections. To see this effect,  we focus on the Planck-scale suppressed
operators that contain both the Higgs $(\Sigma, H, \overline{H})$ and
hidden-sector fields.

Again, we focus on the dimension-five effective operators. For the
K\"{a}hler-type interactions, we have
\begin{align}
 \Delta K_{\rm eff}^{Z} =
\frac{1}{\sqrt{3} M_P}
\Bigl[
\kappa_\Sigma Z |\Sigma|^2 + \kappa_{H} Z |H|^2 + 
\kappa_{\bar{H}} Z |\overline{H}|^2 
+ \kappa^\prime_\Sigma Z^* {\rm Tr}(\Sigma^2) 
+ \kappa^\prime_H Z^* H \overline{H}
+{\rm h.c.}
\Bigr]~,
\label{eq:delkz}
\end{align}
where $Z$ stands for the SUSY-breaking field in the hidden sector. As for
the superpotential operators, in principle, renormalizable interactions
such as $Z{\rm Tr}(\Sigma^2)$ and $ZH\overline{H}$ are allowed, but here
we just assume that the Higgs fields are sequestered from the
SUSY-breaking field so that there is no coupling at the renormalizable
level. In this case, the dominant contribution comes from the
dimension-five operators:
\begin{align}
 \Delta W_{\rm eff}^Z 
=\frac{1}{\sqrt{3} M_P} \Bigl[
\omega_{\lambda} Z\overline{H} \Sigma H 
+\frac{1}{6} \omega_{\lambda^\prime} Z {\rm Tr}\Sigma^3 
+ \omega_\Sigma Z^2 {\rm Tr}\Sigma^2
+ \omega_H Z^2 H\overline{H}
\Bigr]~.
\label{eq:delwz}
\end{align}

Once the SUSY-breaking field $Z$ develops an $F$-term, $F_Z$, the fourth
and fifth terms in Eq.~\eqref{eq:delkz} shift the
$\mu$-terms as
\begin{equation}
 \mu_\Sigma \to \mu_\Sigma + \kappa^\prime_\Sigma m^*_{3/2} ~,
\qquad 
 \mu_H \to \mu_H + \kappa^\prime_H m^*_{3/2} ~,
\end{equation}
where we have used 
\begin{equation}
 m_{3/2} = \frac{F_Z}{\sqrt{3} M_P} ~.
\end{equation}
As was the case for the GM terms, these corrections are smaller than the tree-level terms by a
factor of ${\cal O}(m_{3/2}/M_{\rm GUT})$, and thus can safely be
neglected. The above operators also yield soft SUSY-breaking
terms. A straightforward calculation leads to 
\begin{align}
& \Delta A_{\lambda } = (\kappa_\Sigma + \kappa_H + \kappa_{\bar{H}} +
 \omega_\lambda) m_{3/2} ~,
 \qquad
 \Delta A_{\lambda^\prime} = (3 \kappa_\Sigma + \omega_{\lambda^\prime}) m_{3/2} ~, 
 \nonumber \\[2pt]
& \Delta A_{\bf 10} = \kappa_H m_{3/2} ~,
\qquad 
 \Delta A_{\overline{\bf 5}} = \kappa_{\bar{H}} m_{3/2} ~, 
\nonumber \\[2pt]
 &\Delta B_H = (\kappa_H + \kappa_{\bar{H}}) m_{3/2}~,
\qquad
\Delta B_\Sigma = 2 \kappa_\Sigma m_{3/2} ~,
\nonumber \\[2pt]
& \Delta m_H^2 = |\kappa_H|^2 |m_{3/2}|^2 ~,
\qquad
 \Delta m_{\bar{H}}^2 = |\kappa_{\bar{H}}|^2 |m_{3/2}|^2 ~,
\qquad
\Delta m_\Sigma^2 = |\kappa_\Sigma|^2 |m_{3/2}|^2 ~.
\label{eq:softinduced}
\end{align}
Notice that the last two terms in Eq.~\eqref{eq:delwz} do not generate
soft SUSY-breaking terms.\footnote{Here, we have assumed that the VEV of the lowest
component of the SUSY-breaking field is negligibly small: $|\langle Z
\rangle| \ll M_{\rm GUT}$.}
Hence, we can neglect these operators in what follows.
We also note that if $\kappa_H, \kappa_{\bar{H}} \neq 0$, $A$-terms
for matter fields are generated even though there is no direct
coupling between the matter fields and the SUSY-breaking field. Since
$A$-terms are induced only via quantum corrections in minimal PGM, the corrections in Eq.~\eqref{eq:softinduced} can be
the dominant contribution in this formulation.

When we write down the operators in Eqs.~\eqref{eq:delkz} and
\eqref{eq:delwz}, we have assumed that the field $Z$ is a singlet
field. In this case, $Z$ can in principle appear in the gauge kinetic
function to generate a gaugino mass at tree level, which may dominate the
AMSB contribution \eqref{eq:m5anom}. In the present work,
we instead assume that the Lagrangian is (approximately) invariant under
the shift symmetry\footnote{If the shift symmetry is exact, the
imaginary component of $Z$ becomes massless. This direction can be
stabilized by simply adding a shift-symmetry-violating mass term or by
exploiting strong moduli stabilization as was done in
Ref.~\cite{Evans:2013nka}, with a UV completion discussed in
Ref.~\cite{Kallosh:2006dv}. }
\begin{equation}
 Z \to Z+ i\zeta ~,
\end{equation}
with $\zeta$ real. This can be realized if
\begin{equation}
 \kappa_\Sigma^\prime = \kappa_H^\prime = \omega_\lambda =
  \omega_{\lambda^\prime} =\omega_\Sigma = \omega_H = 0~,
\end{equation}
and $\kappa_\Sigma$, $\kappa_H$, and $\kappa_{\bar{H}}$ are real. In
this case, $Z$ cannot couple to the gauge kinetic function and thus the
dominant contribution to the gaugino mass term is given
by AMSB as in Eq.~\eqref{eq:m5anom}. All of the
superpotential terms \eqref{eq:delwz} are forbidden by the shift
symmetry, {\it i.e.}, $\Delta W_{\rm eff}^Z = 0$, while the K\"{a}hler
potential in Eq.~\eqref{eq:delkz} reduces to 
\begin{equation}
  \Delta K_{\rm eff}^{Z} =
\frac{Z+Z^*}{\sqrt{3} M_P}
\Bigl[
\kappa_\Sigma |\Sigma|^2 + \kappa_{H}  |H|^2 + 
\kappa_{\bar{H}} |\overline{H}|^2 
\Bigr]~.
\label{eq:delkz2}
\end{equation}
We focus on this case unless otherwise noted.

\subsection{GUT-scale matching conditions}
\label{sec:matching}

In this subsection, we summarize the matching conditions at the
unification scale $M_{\rm GUT}$, where the SU(5) GUT parameters are
matched onto the MSSM parameters.

Let us begin with the matching conditions for the gauge coupling
constants. At one-loop level in the $\overline{\rm DR}$ renormalization
scheme \cite{Siegel:1979wq}, we have 
\begin{align}
 \frac{1}{g_1^2(Q)}&=\frac{1}{g_5^2(Q)}+\frac{1}{8\pi^2}\biggl[
\frac{2}{5}
\ln \frac{Q}{M_{H_C}}-10\ln\frac{Q}{M_X}
\biggr]+\frac{8cV}{M_P} (-1)
~, \label{eq:matchg1} \\
 \frac{1}{g_2^2(Q)}&=\frac{1}{g_5^2(Q)}+\frac{1}{8\pi^2}\biggl[
2\ln \frac{Q}{M_\Sigma}-6\ln\frac{Q}{M_X}
\biggr]+\frac{8cV}{M_P} (-3)
~, \\
 \frac{1}{g_3^2(Q)}&=\frac{1}{g_5^2(Q)}+\frac{1}{8\pi^2}\biggl[
\ln \frac{Q}{M_{H_C}}+3\ln \frac{Q}{M_\Sigma}-4\ln\frac{Q}{M_X}
\biggr]+\frac{8cV}{M_P} (2)~,
\end{align}
where $g_1$, $g_2$, and $g_3$ are the U(1)$_Y$, SU(2)$_L$, and SU(3)$_C$ gauge
couplings, respectively, and $Q \simeq M_{\rm GUT}$ is a renormalization
scale. The last terms are induced by the dimension-five operator in
Eq.~\eqref{eq:SigmaWW}. From these equations, we find
\begin{align}
 \frac{3}{g_2^2(Q)} - \frac{2}{g_3^2(Q)} -\frac{1}{g_1^2(Q)}
&=-\frac{3}{10\pi^2} \ln \left(\frac{Q}{M_{H_C}}\right)
-\frac{96cV}{M_P}
~,\label{eq:matchmhc} \\[3pt]
 \frac{5}{g_1^2(Q)} -\frac{3}{g_2^2(Q)} -\frac{2}{g_3^2(Q)}
&= -\frac{3}{2\pi^2}\ln\left(\frac{Q^3}{M_X^2 M_\Sigma}\right) ~,
\label{eq:matchmgut}
\\[3pt]
 \frac{5}{g_1^2(Q)} +\frac{3}{g_2^2(Q)} -\frac{2}{g_3^2(Q)}&= -\frac{15}{2\pi^2} \ln\left(\frac{Q}{M_X}\right) + \frac{6}{g_5^2(Q)} -\frac{144cV}{M_P} ~.\label{eq:matchg5}
\end{align}
Notice that the contribution of the operator \eqref{eq:SigmaWW} to the
right-hand side in Eq.~\eqref{eq:matchmgut} vanishes; we can therefore
unambiguously determine the combination $M_X^2 M_\Sigma$ via
\eqref{eq:matchmgut} by running the gauge couplings up from their
low-energy values~\cite{Hisano:1992mh, Hisano:1992jj,
Hisano:2013cqa}. On the other hand, the operator \eqref{eq:SigmaWW} does
affect the determination of $M_{H_C}$ through
Eq.~\eqref{eq:matchmhc}. Indeed, we can use this degree of freedom to
regard $M_{H_C}$ (or, $\lambda$ and $\lambda^\prime$) as a free
parameter, as was done in Refs.~\cite{azar, eenno}.

For the Yukawa couplings, we use the tree-level matching
conditions. As we mentioned in Sec.~\ref{sec:nrowosbf}, the
higher-dimensional operators in \eqref{eq:weffdelh} affect the matching
conditions, introducing a theoretical uncertainty. 
Without any rationale for evaluating the size of this effect, in this
paper, we simply use 
\begin{equation}
 h_{{\bf 10}, 3} = \frac{1}{4}f_{u_3}~, \qquad
 h_{\overline{\bf 5}, 3} = \frac{f_{d_3} + f_{e_3}}{\sqrt{2}} ~,
\end{equation}
for the third generation Yukawa couplings,
where $h_{{\bf 10}, i}$, $h_{\overline{\bf 5}, i}$, $f_{u_i}$,
$f_{d_i}$, and $f_{e_i}$ are eigenvalues of $h_{\bf 10}$,
$h_{\overline{\bf 5}}$, the up-type Yukawa couplings, the
down-type Yukawa couplings, and the charged lepton Yukawa couplings,
respectively. This condition is the same as that used in
Ref.~\cite{emo}. For the first and second generation Yukawa couplings,
on the other hand, we use
\begin{equation}
 h_{{\bf 10}, i} = \frac{1}{4} f_{u_i} ~, \qquad
 h_{\overline{\bf 5}, i} = \sqrt{2} f_{d_i} ~,
\end{equation}
as in Ref.~\cite{azar}, though the replacement of $f_{d_i}$ by
$f_{e_i}$ or $(f_{d_i}+f_{e_i})/2$ barely changes our result since these
values are very small.

Next we list the matching conditions for the soft SUSY-breaking
terms. For the gaugino masses \cite{Tobe:2003yj, Hisano:1993zu}, we
have\footnote{Notice that we flip the sign of $A$- and $B$-terms from
those used in Refs.~\cite{Tobe:2003yj, Hisano:1993zu} to match the
formulae with our sign convention. }
\begin{align}
 M_1 &= \frac{g_1^2}{g_5^2} M_5
-\frac{g_1^2}{16\pi^2}\left[10 M_5 -10(A_{\lambda^\prime} -B_\Sigma)
 -\frac{2}{5}B_H\right]
-\frac{4cg_1^2V(A_{\lambda^\prime} -B_\Sigma)}{M_P} ~,
\label{eq:m1match}
\\[3pt]
M_2 &= \frac{g_2^2}{g_5^2} M_5
-\frac{g_2^2}{16\pi^2}\left[6 M_5 - 6A_{\lambda^\prime} +4B_\Sigma
 \right]
-\frac{12cg_2^2V(A_{\lambda^\prime} -B_\Sigma)}{M_P} ~,
\label{eq:m2match}
\\[3pt]
M_3 &= \frac{g_3^2}{g_5^2} M_5
-\frac{g_3^2}{16\pi^2}\left[4 M_5 - 4A_{\lambda^\prime} + B_\Sigma
-B_H \right]
+\frac{8cg_3^2V(A_{\lambda^\prime} -B_\Sigma)}{M_P}
~.
\label{eq:m3match}
\end{align}
It is instructive to investigate these equations for the case of
PGM without the higher-dimensional operators ($c =
\kappa_\Sigma = \kappa_H = \kappa_{\bar H} = 0$). By substituting Eqs.~\eqref{eq:aanom}, \eqref{eq:m5anom}, and
\eqref{eq:banom} into the above equations (with $c_\Sigma = c_H = 0$),
at the leading order, we obtain
\begin{align}
 M_1 &\simeq \frac{g_1^2}{16\pi^2} \left(b_5 + 10 -
 \frac{2}{5}\right)m_{3/2}
=  \frac{g_1^2}{16\pi^2} \frac{33}{5} m_{3/2} ~, \nonumber \\
 M_2 &\simeq \frac{g_2^2}{16\pi^2} \left(b_5 + 4 \right)m_{3/2}
=  \frac{g_2^2}{16\pi^2}  m_{3/2} ~, \nonumber \\
 M_3 &\simeq \frac{g_3^2}{16\pi^2} b_5 \, m_{3/2}
= -3 \frac{g_3^2}{16\pi^2}  m_{3/2} ~,
\label{eq:amsbmssm}
\end{align}
which reproduces the gaugino mass spectrum in the MSSM with AMSB 
\cite{anom}. This demonstrates the
well known feature of AMSB, namely, that the contribution of heavy
fields completely decouples at their mass threshold so that the AMSB 
mass spectrum is unambiguously determined in low-energy
effective theories. However, if there exist extra sources for soft masses of
${\cal O}(m_{3/2})$, such as those mediated by the Planck-scale
suppressed operators discussed in Sec.~\ref{sec:nrowsbf}, the gaugino
masses can be modified by an ${\cal O}(1)$ factor and thus the resultant
gaugino mass spectrum can be changed significantly.

The soft masses of the MSSM matter fields, as well as
the $A$-terms of the third generation sfermions, are obtained via the
tree-level matching conditions:
\begin{align}
 m^2_{Q} = m_{U}^2 = m^2_{E} = m^2_{{\bf 10}} ~,&
~~~~~~ m_{D}^2 = m_{L}^2 = m_{\overline{\bf 5}}^2 ~, \nonumber \\
 m_{H_u}^2 = m_H^2 ~,& ~~~~~~ m_{H_d}^2 = m_{\overline{H}}^2 ~,
\nonumber \\
 A_t = A_{\bf 10} ~,& ~~~~~~
 A_b = A_\tau = A_{\overline{\bf 5}} ~.
\end{align}

The $\mu$ and $B$-term matching conditions require fine-tuning such that
the low-energy values of these parameters are ${\cal O}(m_{3/2})$. The
matching conditions are \cite{Borzumati:2009hu}
\begin{align}
 \mu &= \mu_H - 3 \lambda V\left[
1+ \frac{A_{\lambda^\prime} -B_\Sigma}{2 \mu_\Sigma}
\right] ~,
\label{eq:matchingmu}
 \\[3pt]
 B &= B_H + \frac{3\lambda V \Delta}{\mu}
+ \frac{6 \lambda}{\lambda^\prime \mu} \left[
(A_{\lambda^\prime} -B_\Sigma) (2 B_\Sigma -A_{\lambda^\prime}
 +\Delta) -m_\Sigma^2
\right]~,
\label{eq:matchingb}
\end{align}
with
\begin{equation}
 \Delta \equiv A_{\lambda^\prime} - B_\Sigma - A_\lambda +B_H ~.
\label{eq:deltadef}
\end{equation}
As one can see, in order for the left-hand sides of Eqs.~\eqref{eq:matchingmu}
and \eqref{eq:matchingb} to be ${\cal O}(m_{3/2})$, $|\mu_H -3\lambda
V|$ and $|\Delta|$ should be $\lesssim {\cal O}(m_{3/2})$ and ${\cal
O}(m_{3/2}^2/M_{\rm GUT})$, respectively. We accept this fine-tuning in this model.  

Note that the condition $\Delta = 0$ is invariant under the
renormalization-group flow \cite{Kawamura:1994ys}. Therefore, we can
discuss the implications of this condition for the input soft
SUSY-breaking parameters without suffering from the uncertainty due to
renormalization effects. By substituting Eqs.~\eqref{eq:banom} and
\eqref{eq:softinduced} into $\Delta$, we obtain its value at the input
scale \footnote{In CMSSM-like models such as those considered in \cite{azar}, there is no high scale boundary condition for the bilinear terms at $M_{\rm in}$. Therefore, Eq. (\ref{eq:matchingb}) can be used with $\Delta  = 0$ to determine $B_H$ and $B_\Sigma$ at $M_{\rm in}$.  However in a no-scale supergravity model such as that discussed in \cite{eenno} or mSUGRA-like models including PGM, the boundary conditions are $B_0 = A_0 - m_{3/2} \simeq -m_{3/2}$ at $M_{\rm in}$ and 
a non-zero value for $\Delta$ is required to satisfy (\ref{eq:matchingb}). This can be achieved using either a GM term as in Eq. (\ref{eq:banom}) or through the dimension 5 operator dependent on $\omega_\lambda$ and $\omega_\lambda'$.  Note that the GM term in  Eq. (\ref{eq:banom}) carries a factor of two relative to that used in \cite{eenno} which stems from the difference between using twisted matter fields here, relative to untwisted fields in the no-scale construction \cite{eioy2}. }: 
\begin{equation}
 \Delta = \left(\omega_{\lambda^\prime} - \omega_\lambda
\right) \, m_{3/2} + \left(
 \frac{2 c_H}{\mu_H} - \frac{2 c_\Sigma}{\mu_\Sigma}
\right)\, m_{3/2}^2 ~.
\end{equation}
This result indicates that we need an ${\cal O}(m_{3/2}/M_{\rm GUT})$
fine-tuning in $\omega_{\lambda} - \omega_{\lambda^\prime}$ if they are both of 
${\cal O}(1)$. This fine-tuning can, however, be evaded if we consider a
setup as in Sec.~\ref{sec:nrowsbf}, where $\omega_{\lambda} =
\omega_{\lambda^\prime} = 0$ is assured. On the other hand, the second
term is of ${\cal O}(m_{3/2}^2/M_{\rm GUT})$, and thus even if
this term is present the low-energy $B$-parameter remains ${\cal
O}(m_{3/2})$, shifted by
\begin{equation}
 \Delta B = \frac{3\lambda V}{\mu}\left(
 \frac{2 c_H}{\mu_H} - \frac{2 c_\Sigma}{\mu_\Sigma}
\right)\, m_{3/2}^2
= 2 \left(c_H - \frac{12 \lambda c_\Sigma}{\lambda^\prime}\right) \frac{
m_{3/2}^2}{\mu} ~.
\label{eq:deltbgm}
\end{equation}
We will exploit this shift to realize electroweak
symmetry breaking and the matching condition in Eq.~\eqref{eq:matchingb}. 

The $\mu$ and $B$ parameters at low energies are subject to the electroweak
vacuum conditions,
\begin{align}
 \mu^2 &= \frac{m_{1}^2 -m_{2}^2 \tan^2\beta + \frac{1}{2} m_Z^2
 (1-\tan^2\beta ) +\Delta_\mu^{(1)}}{\tan^2 \beta -1 +\Delta_\mu^{(2)}}
 ,\label{eq:muew} \\
 B\mu &= -\frac{1}{2}(m_1^2 + m_2^2 +2 \mu^2) \sin 2\beta +\Delta_B ~,
\label{eq:bew}
\end{align}
where $\Delta_B$ and $\Delta_\mu^{(1,2)}$ denote loop corrections
\cite{Barger:1993gh}. The GUT-scale values of these parameters are 
obtained using renormalization group equations from these values with
some iterations. To satisfy the condition \eqref{eq:matchingb} with the
$B$-parameters in minimal SU(5) at the unification scale, and which are
obtained from given input values via renormalization group
equations, we utilize the Giudice-Masiero contribution to the
$B$-parameter shown in Eq.~\eqref{eq:deltbgm}, as in
Ref.~\cite{eenno}. With the extra free parameters, $c_\Sigma$
and $c_H$, we can always satisfy the matching condition
\eqref{eq:matchingb}.

\subsection{Summary}
\label{sec:modelsummary}

In summary, we consider the minimal SU(5) GUT with PGM where the input scale $M_{\rm in}$ for soft SUSY-breaking
parameters is taken to be higher than the unification scale $M_{\rm
GUT}$. We also introduce GM terms \eqref{eq:gm},
the Planck-scale suppressed correction to the gauge kinetic term
\eqref{eq:SigmaWW}, and the K\"{a}hler-type
non-renormalizable interactions \eqref{eq:delkz2}, neglecting the other
dimension-five operators. The soft
mass parameters at the input scale $M_{\rm in}$ are then given by
\begin{align}
  \left(m_{\bf 10}^2\right)_{ij} &=
\left(m_{\overline{\bf 5}}^2\right)_{ij}
= |m_{3/2}|^2 \, \delta_{ij} ~,
\quad
m_\Sigma^2 = (1+ |\kappa_\Sigma|^2) |m_{3/2}|^2 ~,
\nonumber \\[3pt]
m_H^2 &= (1 + |\kappa_H|^2) |m_{3/2}|^2~, 
\quad m_{\overline{H}}^2 = (1 + |\kappa_{\bar{H}}|^2) |m_{3/2}|^2~,
\nonumber \\[3pt]
A_\lambda &= (\kappa_\Sigma + \kappa_H + \kappa_{\bar{H}})\, m_{3/2} ~,
\quad
A_{\lambda^\prime} = 3 \kappa_\Sigma \, m_{3/2} ~,
\nonumber \\[3pt]
  A_{\bf 10} &= \kappa_H \, m_{3/2} ~,
\quad 
  A_{\overline{\bf 5}} = \kappa_{\bar{H}} \,m_{3/2} ~, 
\nonumber \\[3pt]
 B_H &= (\kappa_H + \kappa_{\bar{H}}-1) m_{3/2}~,
\quad
B_\Sigma = (2 \kappa_\Sigma -1) m_{3/2} ~,
\nonumber \\[3pt]
M_5 &= \frac{b_5 g_5^2}{16\pi^2} m_{3/2} ~.
\end{align}
The $\mu$ and $B$ parameters at the electroweak scale are determined for
a given $\tan \beta$ and $\text{sign}(\mu)$ according to
Eqs.~\eqref{eq:muew} and \eqref{eq:bew}; the values of these parameters
at the unification scale are set by Eqs.~\eqref{eq:matchingmu} and
\eqref{eq:matchingb}, by fine-tuning $\mu_H$ and a linear combination of
$c_H$ and $c_\Sigma$. Our minimal superGUT PGM model
is therefore specified with the following set of input parameters:
\begin{equation}
 m_{3/2}, \quad 
 M_{\rm in}, \quad
 \lambda, \quad
 \lambda^\prime, \quad
 \tan\beta, \quad
 \text{sign}(\mu), \quad
 \kappa_\Sigma, \quad
 \kappa_H, \quad
 \kappa_{\bar{H}}, 
\end{equation}
where $\lambda$ and $\lambda^\prime$ are given at the unification scale
while $\kappa_\Sigma, \kappa_H, \kappa_{\bar{H}}$ are specified at the
input scale $M_{\rm in}$. The coefficient of the operator
\eqref{eq:SigmaWW}, $c$, as well as the unified coupling, $g_5$, and the
adjoint Higgs VEV, $V$, are determined using
Eqs.~(\ref{eq:matchmhc}--\ref{eq:matchg5}) as was done in
Refs.~\cite{azar, eenno}. In the following analysis,
however, we also consider the case where $c = 0$; in this case,
$\lambda^\prime$ becomes a dependent parameter and is determined by the
conditions (\ref{eq:matchmhc}--\ref{eq:matchg5}). 
Our calculations require that we choose a value of $Q$ to evaluate the gauge couplings and implement the matching conditions. We have chosen $\log_{10} (Q/{\rm 1~GeV})  = 16.3$ throughout this work, and our results
are relatively insensitive to this choice (so long as it is near $10^{16}$ GeV).

\section{Proton decay}
\label{sec:protondecay}

In the minimal SUSY GUT, the baryon- and lepton-number violating
dimension-five operators are generated by the exchange of the
color-triplet Higgs fields $H_C$ and $\overline{H}_C$. Such operators are
potentially quite dangerous due to their low dimensionality, since they
may induce rapid proton decay. In fact, it was argued in
Ref.~\cite{Murayama:2001ur} based on the calculation of
Ref.~\cite{Goto:1998qg} that if stops lie around the TeV scale, the
proton decay lifetime induced by the color-triplet Higgs exchange is too
short to evade the experimental limit. This problem is, however, evaded
if the SUSY scale is much higher than the TeV scale
\cite{Hisano:2013exa, McKeen:2013dma, Nagata:2013sba, Nagata:2013ive,
Evans:2015bxa, eelnos}, as in PGM
models. Indeed, it is shown in Ref.~\cite{Evans:2015bxa} that in
minimal SU(5) with PGM, the lifetime of the $p \to K^+
\bar{\nu}$ mode is much longer than the current experimental bound for
$m_{3/2} \gtrsim 100$~TeV. It is found that this is also true for the
present model and thus we can safely neglect dimension-five proton
decay.

When dimension-five proton decay is suppressed,  dimension-six
proton decay induced by the exchange of GUT gauge bosons becomes
dominant. The primary decay channel is $p \to  \pi^0 e^+$ and its rate goes as
$\propto M_X^{-4}$. Intriguingly, in the models considered, $M_X$ can be
predicted to be lower than the typical GUT scale for TeV-scale
SUSY models, $\simeq 2 \times 10^{16}$~GeV, which then results in a
shorter proton decay lifetime. As discussed in
Ref.~\cite{Hisano:2013cqa}, the combination
$(M_X^2 M_\Sigma)^{1/3}$, scales as $(M_3 M_2)^{-1/9}$. Hence, $M_X$
tends to decrease when $m_{3/2}$ is large.\footnote{This tendency was also found
in Ref.~\cite{Evans:2015bxa}.} 
In addition, as we discuss below, since $M_\Sigma \sim \lambda' V$ a small value for $M_X$ can also be obtained when
$\lambda^\prime$ is ${\cal O}(1)$.
Motivated by these observations, we study the lifetime of the $p
\to  \pi^0 e^+$ decay channel predicted in our model in this paper. We
review the calculation of the dimension-six proton decay in
this section, and show our numerical results in the subsequent section. 

As we mentioned above, dimension-six proton decay is induced by
SU(5) gauge interactions. The relevant Lagrangian terms are 
\begin{align}
 {\cal L}_{\rm int} = \frac{g_5}{\sqrt{2}}
\left[
- \overline{d^c_{Ri}} \Slash{X} L_i
+ e^{-i\varphi_i}\overline{Q}_i \Slash{X} u_{Ri}^c
+ \overline{e_{Ri}^c} \Slash{X} (V_{\rm CKM}^\dagger)_{ij}Q_j
+ {\rm h.c.}
\right]~,
\end{align}
where $X$ stands for the SU(5) massive gauge fields, $V_{\rm CKM}$ denotes the
Cabibbo-Kobayashi-Maskawa (CKM) matrix, $\varphi_i$ are the GUT phase
factors \cite{Ellis:1979hy}, and we have suppressed the SU(3)$_C$ and
SU(2)$_L$ indices just for simplicity (see, {\it e.g.},
Ref.~\cite{Hisano:2012wq} for a more explicit expression). We integrate
out the $X$ bosons at tree level to obtain the dimension-six
effective operators, and evolve their Wilson coefficients from the
unification scale to the hadronic scale according to renormalization
group equations in each effective theory. By using these effective
operators, we can then calculate the partial decay width of the $p \to
 \pi^0 e^+$ mode. The result is
\begin{equation}
 \Gamma (p\to  \pi^0 e^+)=
\frac{m_p}{32\pi}\biggl(1-\frac{m_\pi^2}{m_p^2}\biggr)^2
\bigl[
\vert {\cal A}_L(p\to \pi^0 e^+) \vert^2+
\vert {\cal A}_R(p\to \pi^0 e^+) \vert^2
\bigr]~,
\end{equation}
where $m_p$ and $m_\pi$ denote the proton and pion masses,
respectively. The amplitudes ${\cal A}_{L/R}$ are given by
\begin{align}
 {\cal A}_L(p\to \pi^0 e^+)&=
- \frac{g_5^2}{M_X^2} \cdot
A_1 \cdot \langle \pi^0\vert (ud)_Ru_L\vert p\rangle
~,\nonumber \\
 {\cal A}_R(p\to \pi^0 e^+)&=
- \frac{g_5^2}{M_X^2} (1+|V_{ud}|^2) \cdot
A_2 \cdot \langle \pi^0\vert (ud)_Ru_L\vert p\rangle
~,
\end{align}
where $V_{ud} \equiv (V_{\rm CKM})_{12}$ and $\langle \pi^0\vert (ud)_Ru_L\vert p\rangle$ is the hadron matrix
element, for which we use the result obtained by a lattice QCD
simulation \cite{Aoki:2017puj}: 
\begin{align}
\langle \pi^0\vert (ud)_Ru_L\vert p\rangle &= 
- 0.131(4)(13) ~\text{GeV}^2
~.
\end{align}
$A_1$ and $A_2$ are the renormalization factors for the effective
operators. They are given by
\begin{align}
 A_1 &=
A_L \cdot \biggl[
\frac{\alpha_3(M_{\text{SUSY}})}{\alpha_3(M_{\rm GUT})}
\biggr]^{\frac{4}{9}}
\biggl[
\frac{\alpha_2(M_{\text{SUSY}})}{\alpha_2(M_{\rm GUT})}
\biggr]^{-\frac{3}{2}}
\biggl[
\frac{\alpha_1(M_{\text{SUSY}})}{\alpha_1(M_{\rm GUT})}
\biggr]^{-\frac{1}{18}}
\nonumber \\[2pt]
&\times
\biggl[
\frac{\alpha_3(m_Z)}{\alpha_3(M_{\rm SUSY})}
\biggr]^{\frac{2}{7}}
\biggl[
\frac{\alpha_2(m_Z)}{\alpha_2(M_{\rm SUSY})}
\biggr]^{\frac{27}{38}}
\biggl[
\frac{\alpha_1(m_Z)}{\alpha_1(M_{\rm SUSY})}
\biggr]^{-\frac{11}{82}} ~, \nonumber \\[3pt]
 A_2 &=
A_L \cdot \biggl[
\frac{\alpha_3(M_{\text{SUSY}})}{\alpha_3(M_{\rm GUT})}
\biggr]^{\frac{4}{9}}
\biggl[
\frac{\alpha_2(M_{\text{SUSY}})}{\alpha_2(M_{\rm GUT})}
\biggr]^{-\frac{3}{2}}
\biggl[
\frac{\alpha_1(M_{\text{SUSY}})}{\alpha_1(M_{\rm GUT})}
\biggr]^{-\frac{23}{198}}
\nonumber \\[2pt]
&\times
\biggl[
\frac{\alpha_3(m_Z)}{\alpha_3(M_{\rm SUSY})}
\biggr]^{\frac{2}{7}}
\biggl[
\frac{\alpha_2(m_Z)}{\alpha_2(M_{\rm SUSY})}
\biggr]^{\frac{27}{38}}
\biggl[
\frac{\alpha_1(m_Z)}{\alpha_1(M_{\rm SUSY})}
\biggr]^{-\frac{23}{82}} ~,
\end{align}
where $M_{\rm SUSY}$ indicates the SUSY-breaking scale, and $A_L \simeq
1.247$ is the QCD renormalization factor that is computed using the
two-loop renormalization group equation \cite{Nihei:1994tx}. The other
renormalization factors are obtained at the one-loop level by using the
renormalization group equations given in Refs.~\cite{Abbott:1980zj,
Munoz:1986kq}.\footnote{Above the SUSY-breaking scale, the two-loop
renormalization group equations are also available
\cite{Hisano:2013ege}. Moreover, the one-loop threshold corrections at
the GUT scale are computed in Ref.~\cite{Hisano:2015ala}. These
corrections are found to be fairly small, so we neglect them in our
calculation. }

Note that in contrast to dimension-five decay modes, the partial decay width $\Gamma (p \to \pi^0 e^+)$ does not
depend on the GUT phases $\varphi_i$ (see, {\it
e.g.}, Ref.~\cite{azar}). In addition, $g_5$ and $M_X$, which
appear in the amplitudes, can also be evaluated by using the threshold
corrections (\ref{eq:matchmhc}--\ref{eq:matchg5}) as we discussed in the
previous section. For these reasons, the evaluation of the dimension-six proton
decay rate is rather robust and suffers less from unknown
uncertainties.

\section{Results}
\label{sec:result}

We now show our numerical results for the study of the superGUT PGM
model. In Sec.~\ref{sec:wo}, we first discuss the case
without any of the higher-dimensional operators. Then, in Sec.~\ref{sec:wc}, we
consider models which include the operator \eqref{eq:SigmaWW} ($c \neq 0$) but
without the K\"{a}hler-type operators \eqref{eq:delkz2}. Finally, in
Sec.~\ref{sec:all}, we analyze the model with all of the higher-dimensional
operators we consider in this paper.

\subsection{Models without higher-dimensional operators}
\label{sec:wo}

We begin with the case with $c = \kappa_\Sigma = \kappa_{H} =
\kappa_{\bar{H}} = 0$. As we mentioned in Sec.~\ref{sec:modelsummary},
in this case, $\lambda^\prime$ is not an independent parameter,
but rather is fixed by the matching conditions
(\ref{eq:matchmhc}--\ref{eq:matchg5}). Therefore, each parameter point
is specified by $m_{3/2}$, $M_{\rm in}$, $\lambda$, $\tan\beta$, and
$\text{sign}(\mu)$.

\begin{figure}[ht!]
  \centering
  \vskip -1in
  \includegraphics[width=.58\columnwidth]{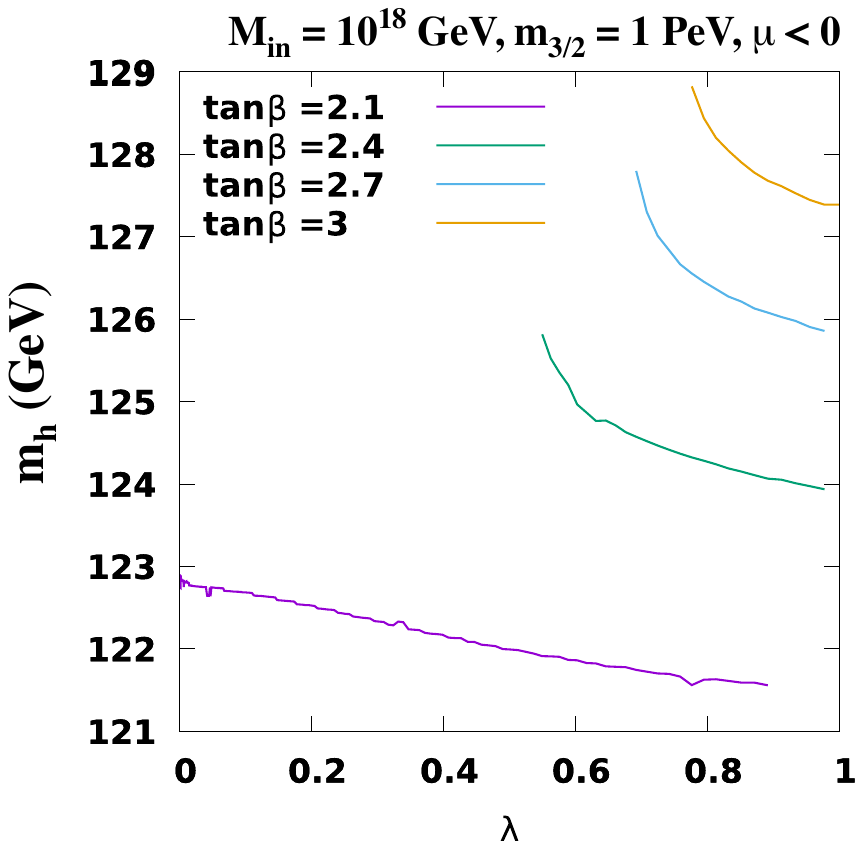} 
  \hskip -1.3in 
\includegraphics[width=0.58\columnwidth]{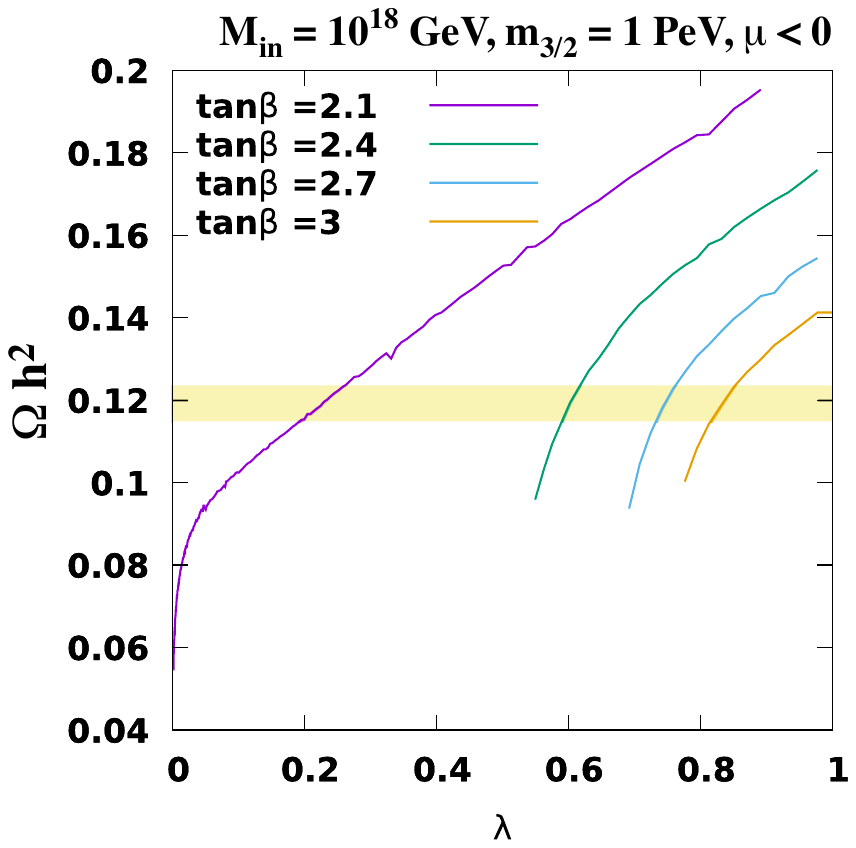}
\\
\vskip -2in
  \includegraphics[width=0.58\columnwidth]{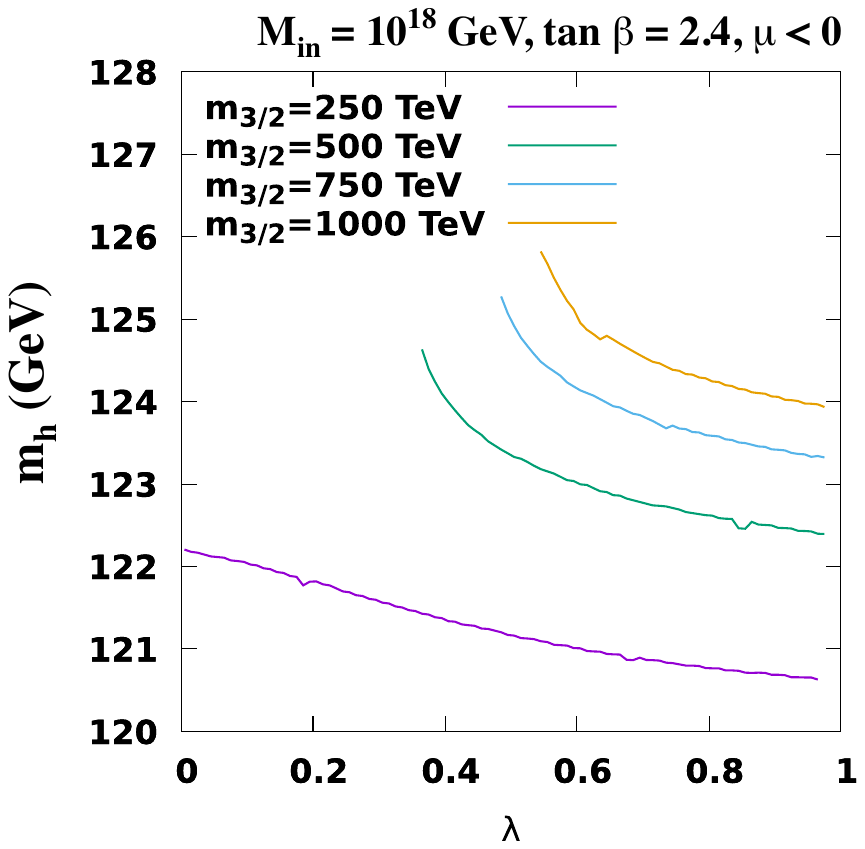}
  \hskip -1.3in 
  \includegraphics[width=0.58\columnwidth]{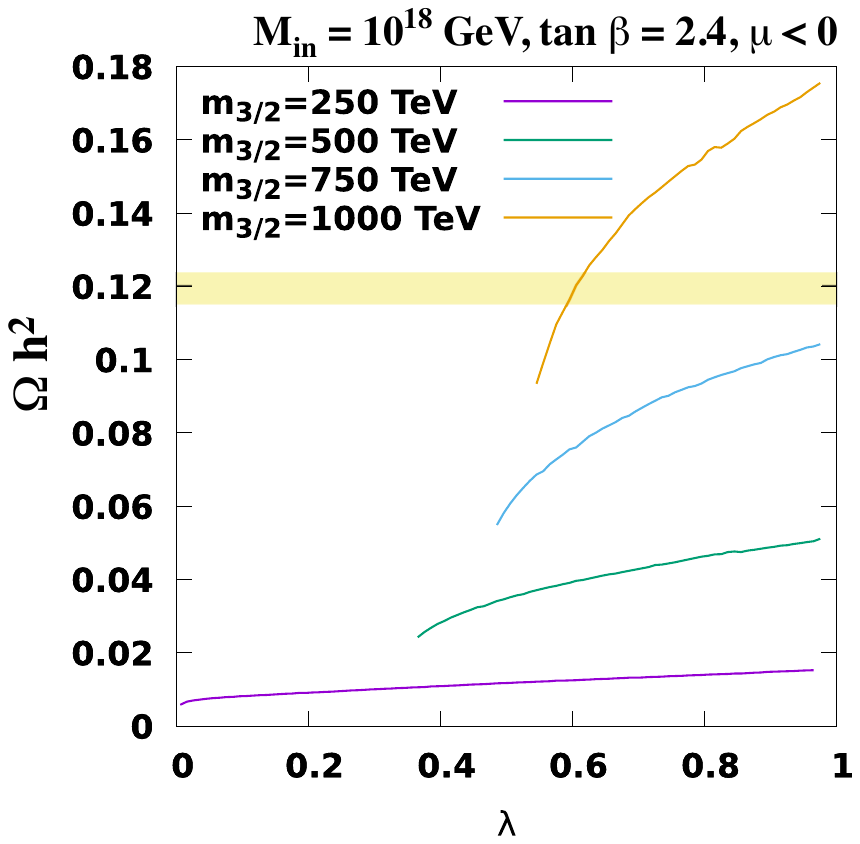}
\vskip -1.in
\caption{\it (a,c) The Higgs mass $m_h$ and (b,d) the LSP density $\Omega_{\rm
 LSP} h^2$ as functions of the coupling $\lambda$ with
 $m_{3/2} = 10^3$~TeV for several values
 of $\tan\beta$ (a,b) or fixed $\tan \beta = 2.4$ for several values of $m_{3/2}$ (c,d).  
 We set $M_{\rm in} =
 10^{18}$~GeV, and $\mu < 0$.
} 
  \label{fig:lamGnoe}
\end{figure}

To maximize the effect of the renormalization group evolution above
$M_{\rm GUT}$, we set $M_{\rm in} = 10^{18}$~GeV and choose $\mu < 0$ so as to maximize the wino mass
for the low values of $\tan \beta$ we are considering\footnote{We include supersymmetric one loop corrections to the anomaly mediated masses \cite{pp} which appear as significant threshold corrections
when the heavy Higgses are integrated out \cite{ggw}. 
These are dependent on the sign of $\mu$.}.
In Fig.~\ref{fig:lamGnoe}a, we plot the predicted value
of the SM-like Higgs mass $m_h$ as a function of the coupling $\lambda$
for several values of $\tan\beta$. 
In order to obtain rEWSB,
PGM models are generally restricted to low values of $\tan \beta$ \cite{eioy}. In addition, unless $\tan \beta$ is near its lower limit of $\sim 2$, the value of $\lambda$ must be relatively large (of order 1). In the Figure, for the three larger values of $\tan \beta$, there is a lower limit to $\lambda$ which allows rEWSB. Very high values of $\lambda$ are also excluded as eventually the running value of $\lambda$ becomes non-perturbative as it runs from $M_{\rm GUT}$ to $M_{\rm in}$. This effect cuts off the curves as $\lambda (M_{\rm GUT})$
approaches 1. 
As we see, $m_h \simeq 125$~GeV is best obtained for $\tan \beta \approx 2.4$. Given that the Higgs-mass
calculation suffers from the uncertainty of a few GeV,\footnote{See
Ref.~\cite{ehow++} for a recent discussion on the uncertainty
in the Higgs mass calculation. } we conclude that the observed value of
$m_h$ can be reproduced for $\tan\beta \simeq 2$--3. 
A similar pattern is seen in Fig.~\ref{fig:lamGnoe}c which shows the Higgs mass for fixed $\tan \beta = 2.4$ for several values of $m_{3/2}$ between 250--1000~TeV.

Since we do not include the higher-dimensional operators
here, the gaugino masses are given by the AMSB spectrum (plus loop corrections),
and thus the LSP is wino-like. It is known that
the thermal relic abundance of the wino-like LSP agrees with the observed
dark matter density $\Omega_{\rm DM} h^2 \simeq 0.12$ \cite{planck18} if
its mass is $\simeq 3$~TeV \cite{Hisano:2006nn}. From
Eq.~\eqref{eq:amsbmssm}, we find that a 3-TeV wino mass can be
obtained for $m_{3/2} \simeq 10^3$~TeV, and thus we expect that the
correct DM density can be reproduced with this gravitino mass. 
In Fig.~\ref{fig:lamGnoe}b, we show the thermal
relic abundance of the LSP as a function of
$\lambda$ with the same choice of parameters used in
Fig.~\ref{fig:lamGnoe}a. It is in fact found that
$\Omega_{\rm LSP} h^2\simeq 0.12$ can be obtained, again for an $\lambda \sim {\cal O}(1)$ and a small value of $\tan \beta$. 
While lowering $m_{3/2}$ to 500~TeV would still allow for 
a Higgs mass of 125~GeV (albeit with a slightly lower value of $\lambda$), the relic density would drop to $\Omega_{\rm LSP} h^2 < 0.05$ due
to the decrease in the wino mass. 
We clearly see the dependence of $\Omega_{\rm LSP} h^2$ on $m_{3/2}$ in Fig.~\ref{fig:lamGnoe}d
where $\tan \beta = 2.4$.

\begin{figure}
  \centering
  \vskip -1in
  \includegraphics[width=.65\columnwidth]{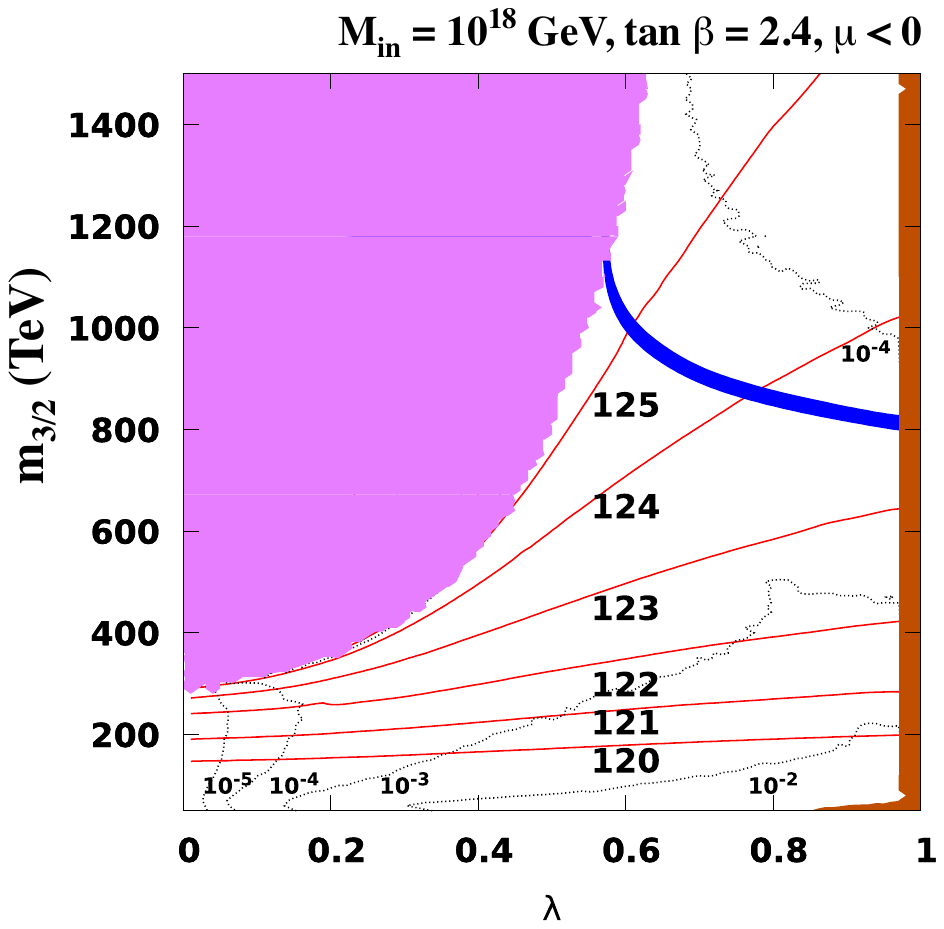}
  \vskip -1.8in
  \includegraphics[width=.65\columnwidth]{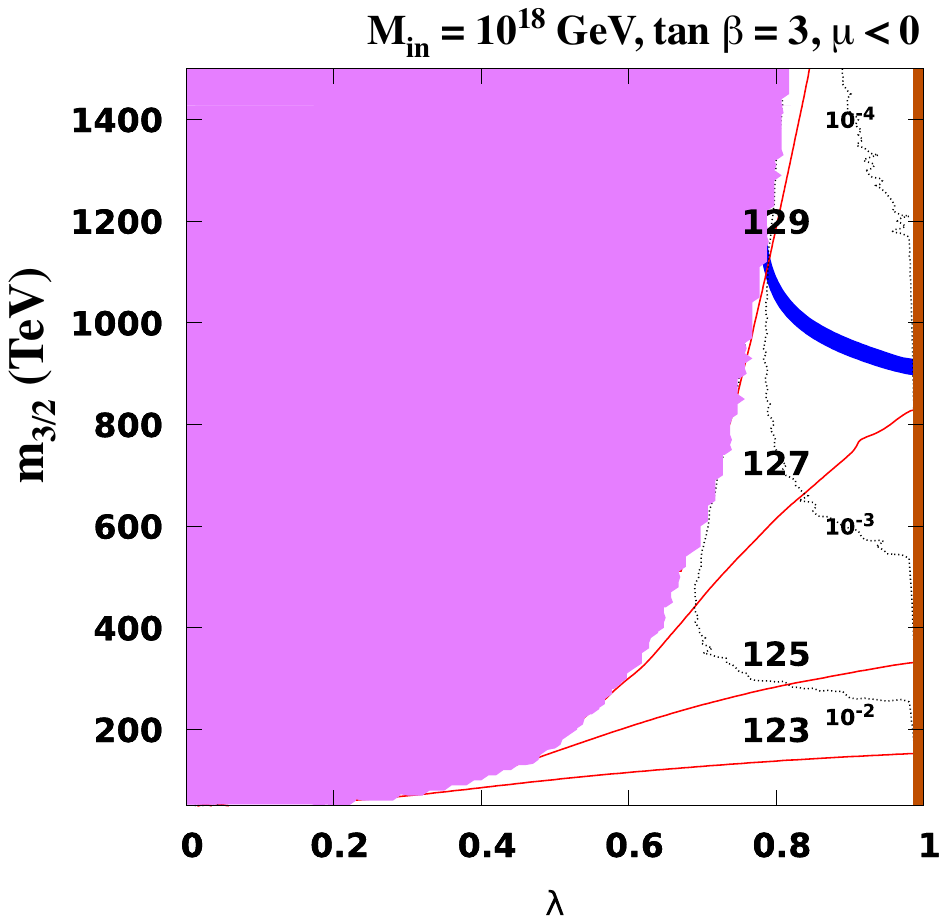}
  \vskip -1in
\caption{\it Examples of the $\lambda$-$m_{3/2}$ plane for $M_{\rm in} = 10^{18}$~GeV, and $\mu < 0$. In the upper panel, $\tan \beta = 2.4$ and in the lower panel,
$\tan \beta = 3$,
The solid red and dotted black curves show the Higgs mass $m_h$ and $\lambda^\prime$, respectively. 
The blue shaded region corresponds to the areas where the LSP abundance agrees with the observed dark matter density. In the magenta region, rEWSB can not be obtained, while in the brown region $\lambda$ becomes non-perturbative at the input scale.
} 
  \label{fig:SuperPGMnoe_1_24_18Con}
\end{figure}

In Fig.~\ref{fig:SuperPGMnoe_1_24_18Con}, we show two examples of $\lambda$-$m_{3/2}$ plane for $\tan \beta = 2.4$ (upper panel), and $\tan \beta = 3$ (lower panel) and in both cases $M_{\rm in} = 10^{18}$~GeV, and $\mu < 0$. The solid red and dotted black curves show the Higgs mass $m_h$ and $\lambda^\prime$, respectively. Recall that with $c=0$, $\lambda'$ is not a free parameter and is determined by the matching conditions. As one can see, when $\tan \beta = 2.4$
over much of the plane, $\lambda'$ takes values between $10^{-4}$ and $10^{-3}$
with slightly more variation when $\tan \beta = 3$. 
The blue shaded region corresponds to the areas where the LSP abundance agrees with the observed DM density, $\Omega_{\rm DM} h^2 \simeq 0.1200 \pm 0.0036$ 
(3$\sigma$)  \cite{planck18}. In the magenta region, rEWSB can not be obtained, while in the brown region $\lambda$ becomes non-perturbative at the input scale (recall that the value used in the figure is $\lambda(M_{\rm GUT})$). This figure shows that we can obtain the correct value of the SM-like Higgs boson mass and the DM abundance simultaneously for $m_{3/2} \sim 1$~PeV and $\lambda = {\cal O}(1)$, which remains perturbative up to the input scale when $\tan \beta \approx 2.4$.  As one can see,  in the lower panel 
when $\tan \beta = 3$ the Higgs mass increases significantly. However, some of the dark matter strip may be viable considering the uncertainty in the Higgs mass calculation. In addition, in this case, the allowed range in $\lambda$ is limited by the 
requirement of rEWSB. 

The relatively large region of parameter space with 
$m_h \sim 125$ GeV and $\Omega_{\rm LSP} h^2 \sim 0.12$
is achieved with a high supersymmetry breaking input
scale, $M_{\rm in} = 10^{18}$ GeV.  Much of this space
disappears as $M_{\rm in}$ approaches the GUT scale.
For example, when $M_{\rm in}$ is lowered to $10^{16.5}$ GeV.
the relic density strip remains near $m_{3/2} \sim 1$ PeV
since that is required to obtain a wino with mass $\sim 3$ TeV.  However, the Higgs mass contours are pushed to lower values of $m_{3/2}$ and more importantly the
region with viable rEWSB is also pushed to lower $m_{3/2}$ so that obtaining the correct Higgs mass
and relic density is no longer possible.

Because $\lambda'$  turns out to be small in this class of models with $c=0$, $M_X$ tends to be quite large. Note that by using Eqs.~\eqref{eq:mx} and \eqref{eq:msigma} we can
express the mass of the SU(5) gauge bosons as 
\begin{equation}
 M_X = \biggl(\frac{2g_5}{\lambda^\prime}\biggr)^{\frac{1}{3}} 
\, \left(M_X^2 M_\Sigma\right)^{\frac{1}{3}} ~.
\label{eq:mxvslamp}
\end{equation}
The factor $\left(M_X^2 M_\Sigma\right)^{\frac{1}{3}}$ can be evaluated
by using Eq.~\eqref{eq:matchmgut}, which is found to have a very small
dependence on the choice of parameters (see, {\it e.g.},
Ref.~\cite{Evans:2015bxa}). As a result, when $\lambda^\prime$ is very
small, the proton decay lifetime becomes
very long. We find that over the parameter space considered in
the above figures the predicted value of the lifetime of the $p
\to \pi^0 e^+$ decay channel is well above the current experimental
bound, $\tau (p \to \pi^0 e^+) > 1.6\times 10^{34}$~years
\cite{Miura:2016krn}.

In summary, we find that despite the limited number of degrees of
freedom, there is a parameter set with which both the Higgs mass and
the dark matter relic abundance can be explained and the predicted value
of proton lifetime is consistent with the experimental bound. The LSP is
a wino-like neutralino as in AMSB models. In the following
subsections, we consider the cases with the Planck-scale suppressed
non-renormalizable operators and discuss their phenomenological
consequences.

\subsection{Models with Superpotential non-renormalizable interactions}
\label{sec:wc}

Next, we discuss the case with $c \neq 0$, keeping $\kappa_\Sigma =
\kappa_H = \kappa_{\bar{H}} = 0$. In this case, we can regard
$\lambda^\prime$ as a free parameter, with $c$ determined through the
matching conditions (\ref{eq:matchmhc}--\ref{eq:matchg5}). A non-zero
value of $c$ has two important phenomenological consequences. First,
since $\lambda^\prime$ is a free parameter, we can take it to be
significantly larger than the values displayed in Fig. \ref{fig:SuperPGMnoe_1_24_18Con} where $c=0$. As can be seen from
Eq.~\eqref{eq:mxvslamp}, a large $\lambda^\prime$ results in a small
$M_X$, which increases the proton decay rate, making proton decay
potentially observable. Second, a non-zero value
of $c$ modifies the gaugino masses through the matching conditions
(\ref{eq:m1match}--\ref{eq:m3match}), which significantly affects the
thermal relic abundance of the LSP. We will examine both of these 
effects in this subsection. 

\begin{figure}[ht]
  \centering
  \vskip -1in
  \includegraphics[width=0.58\columnwidth]{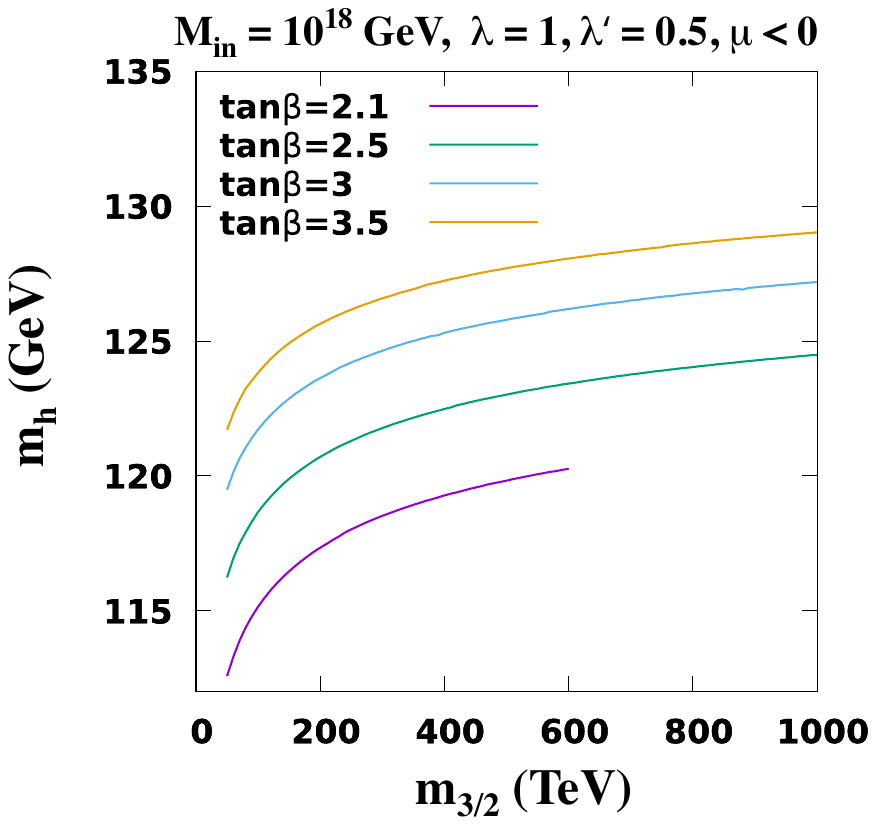}
    \hskip -1.3in 
  \includegraphics[width=0.58\columnwidth]{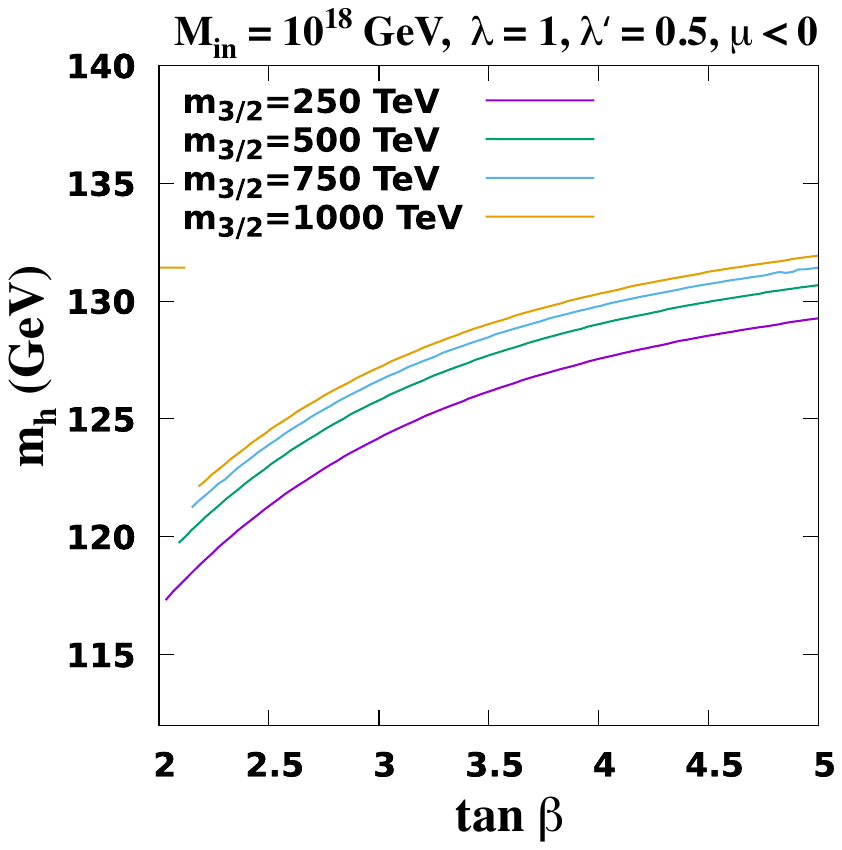}
    \vskip -1in
\caption{\it The Higgs mass $m_h$ as a function of (a)
$m_{3/2}$  for several values of $\tan\beta$ and (b)  as a function of $\tan \beta$ for several values  $m_{3/2}$ with $\lambda =1.0$, $\lambda' = 0.5$, $M_{\rm in} = 10^{18}~\text{GeV}$, and $\mu < 0$ in both panels.} 
  \label{fig:lampG1}
\end{figure}

First, in Fig.~\ref{fig:lampG1}a, we plot the Higgs mass
$m_h$ as a function of $m_{3/2}$ for $\mu < 0$, $\lambda = 1$, $\lambda' = 0.5$ and $M_{\rm
in} = 10^{18}$~GeV, for several values of $\tan\beta$. 
In the right panel (Fig. \ref{fig:lampG1}b),
the Higgs mass is shown as a function of $\tan \beta$ for several values of $m_{3/2}$. The predicted value of
$m_h$ is almost independent of $\lambda^\prime$; $m_h$ mainly depends on
$m_{3/2}$ and $\tan\beta$ as in the minimal PGM model, and thus we can
always explain the observed value $m_h \simeq 125$~GeV by choosing these
two parameters appropriately. 
Higher $m_{3/2}$ is necessary if $\tan \beta < 3$. Recall that $\tan \beta$ much larger than 3 is problematic without large $\lambda$ because 
we lose the ability to achieve rEWSB.

\begin{figure}[ht]
  \centering
  \vskip -1in
  \includegraphics[width=0.58\columnwidth]{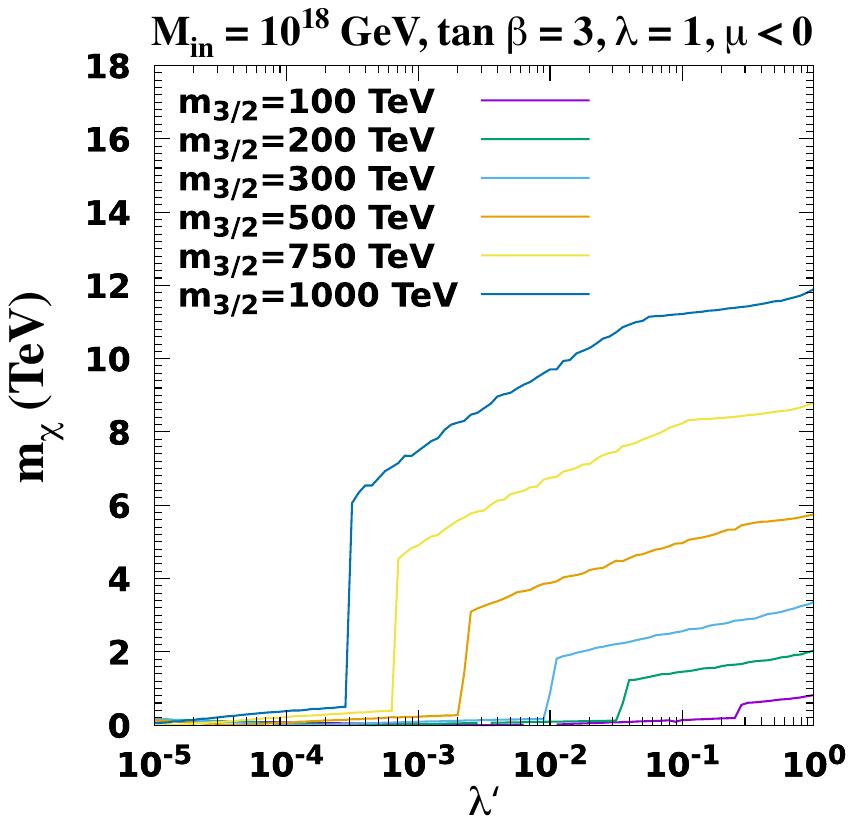}
      \hskip -1.2in 
  \includegraphics[width=0.58\columnwidth]{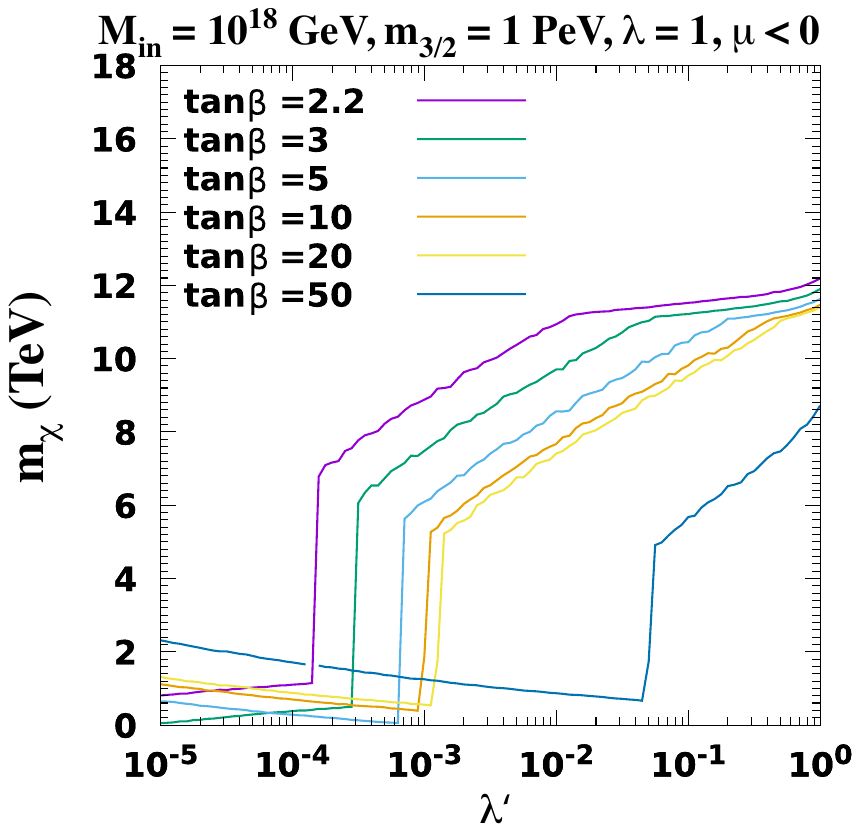}
    \vskip -1in
\caption{\it The mass of the lightest gaugino as a function of $\lambda'$.
In both panels, $M_{\rm in} = 10^{18}$ GeV and $\lambda = 1$ 
with $\mu < 0$. In (a),
$\tan \beta = 3$ showing the gaugino mass for several values of $m_{3/2}$.
In (b), $m_{3/2} = 1$ PeV, for several values of $\tan \beta$.
} 
  \label{fig:lampG2}
\end{figure}

\begin{figure}[ht!]
  \centering
    \vskip -1in
  \includegraphics[width=0.58\columnwidth]{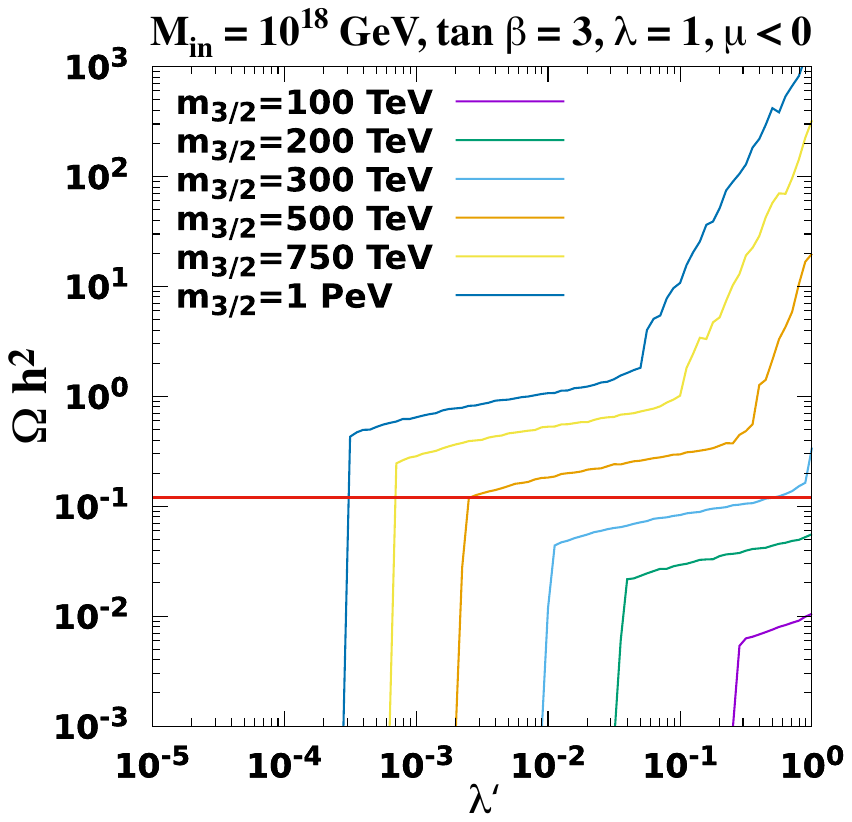}
      \hskip -1.2in 
  \includegraphics[width=0.58\columnwidth]{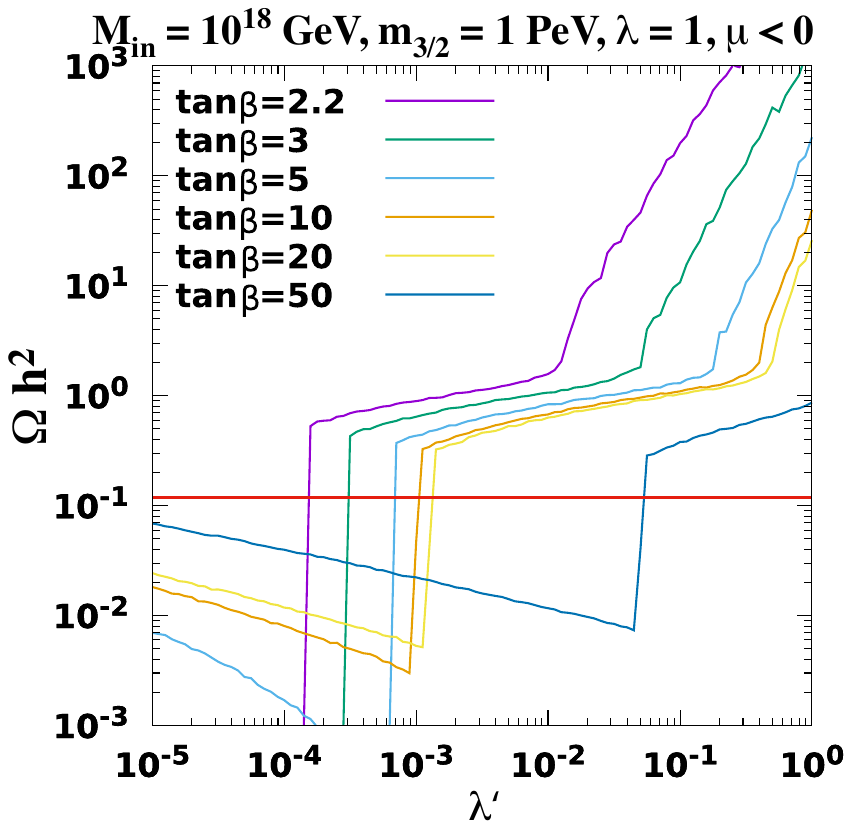}
  \vskip -1in
\caption{\it As in Fig.~\ref{fig:lampG2}, showing the relic density of the lightest gaugino given by $\Omega_{\rm LSP} h^2$ as a function of $\lambda'$.  The red horizontal line shows the Planck value for the relic density.
} 
  \label{fig:lampG3}
\end{figure}

The mass of the LSP (lightest gaugino) is shown in 
Fig.~\ref{fig:lampG2} as a function of 
$\lambda'$, for the same fixed values of $\lambda, M_{\rm in}$,
and sign of $\mu$ used in Fig.~\ref{fig:lampG1}.  
In the left panel, we fix $\tan \beta = 3$ for several choices of $m_{3/2}$ 
whereas in the right panel, we fix $m_{3/2} = 1$ PeV for several choices of $\tan \beta$. 
As can be seen from Eqs.~(\ref{eq:m1match}--\ref{eq:m3match}), the
gaugino masses deviate from the AMSB spectrum if $c \neq 0$.
Depending on the value of $m_{3/2}$ and $\tan \beta$,
there is a critical value for $\lambda'$ such that the 
mass of the lightest gaugino suddenly jumps from
a value $\lesssim 1$ TeV to several TeV.
Recall that $c$ is calculated from the matching conditions
of the gauge couplings which depend on $\lambda^\prime$ through $M_\Sigma$. 
The jump in the gaugino mass seen in Fig. \ref{fig:lampG2} is due to a change in sign in $c$.
Since $c$ affects the gaugino masses
as well, when $c$ changes sign, we see a rapid increase
in the wino mass. 
An immediate consequence of this transition
is seen in Fig. \ref{fig:lampG3}.  For low  $\lambda'$,
the relic density is low as one would expect for a
relatively light wino. At large $\lambda'$,
the mass of the wino increases (due to the change in $c$) to 
several TeV and therefore we see (as expected) an overdensity in the LSP.  
When the sign change in $c$ occurs, it is possible to obtain the correct relic density.  The red horizontal line shows the value of the 
relic density determined by Planck \cite{planck18} and allows one to 
read off the requisite value of $\lambda^\prime$.
Recall that at low $\lambda'$, it is possible
to obtain the correct relic density for lower
values of $\tan \beta$ as we saw in the previous subsection.

\begin{figure}[ht!]
  \centering
  \vskip -1in
  \includegraphics[width=0.58\columnwidth]{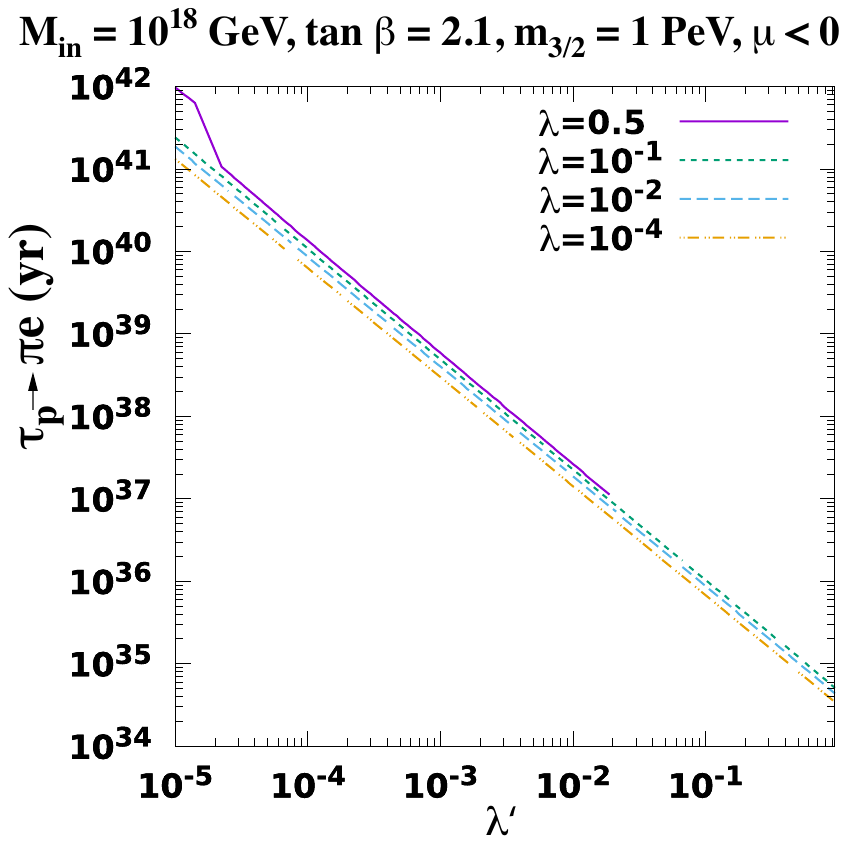}
        \hskip -1.2in 
  \includegraphics[width=0.58\columnwidth]{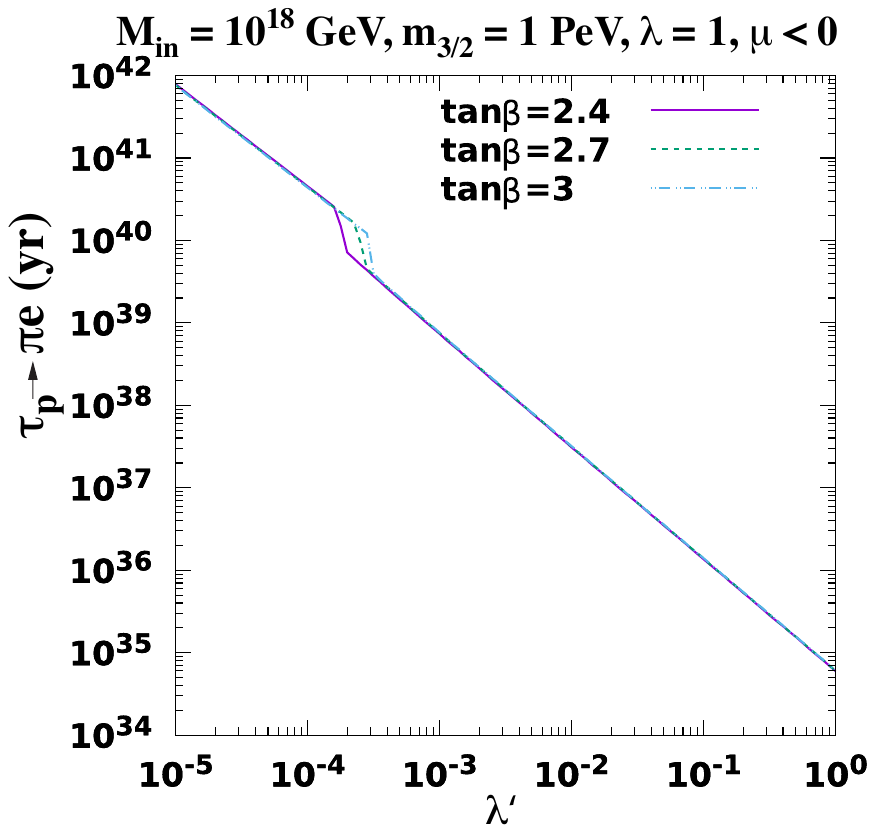}
  \vskip -1in
\caption{\it The lifetime of the $p \to \pi^0 e^+$ channel as a 
function of the coupling $\lambda^\prime$ with $M_{\rm in} = 10^{18}$ GeV,
$m_{3/2} = 1$ PeV, and $\mu < 0$. In (a), $\tan \beta = 2.1$ and 
the lifetime is shown for 
several values of $\lambda$. In (b), $\lambda = 1$ for 
several values of $\tan \beta$.
} 
  \label{fig:lampG4}
\end{figure}

Next, we show in Fig.~\ref{fig:lampG4} the
lifetime of the $p \to \pi^0 e^+$ decay channel as a function of
$\lambda^\prime$ for fixed $M_{\rm in} = 10^{18}$ GeV, $m_{3/2} = 1$ PeV for $\mu < 0$. In the left panel (Fig.~\ref{fig:lampG4}a), we fix $\tan \beta = 2.1$ and show the lifetime for several values of $\lambda$. As we expect, the lifetime decreases as $\lambda^\prime$
increases. The predicted lifetime is always above the current
experimental bound, $\tau (p \to \pi^0 e^+) > 1.6\times 10^{34}$~years
\cite{Miura:2016krn}, but can be within the sensitivity of the
Hyper-Kamiokande, $\tau (p \to \pi^0 e^+) \simeq 7.8 \times
10^{34}$~years \cite{Abe:2018uyc}, if $\lambda^\prime$ is ${\cal O}(1)$. 
The shift in the line at $\lambda' \simeq 10^{-4.7}$ is due
to a sign change in the calculated value of $c$ which affects the gaugino masses and gauge couplings.
In the right panel (Fig. \ref{fig:lampG4}b), we show the lifetime for fixed $\lambda = 1$ for several choices of $\tan \beta$. For this choice of $\lambda = 1$ the discontinuity in the curves due to the sign change of $c$ occurs at higher
$\lambda' \simeq 10^{-3.7}$.

\begin{figure}[ht!]
  \centering
  \vskip -1.2in
  \includegraphics[width=0.65\columnwidth]{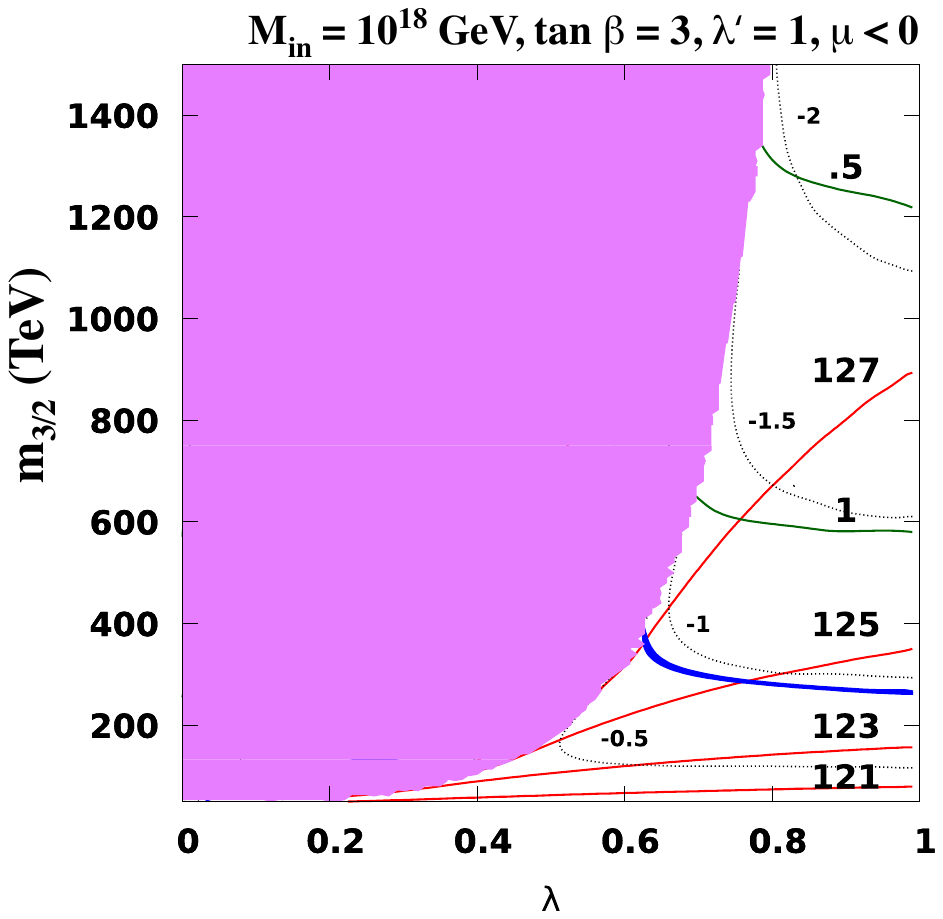}
  \vskip -2.2in
  \includegraphics[width=0.65\columnwidth]{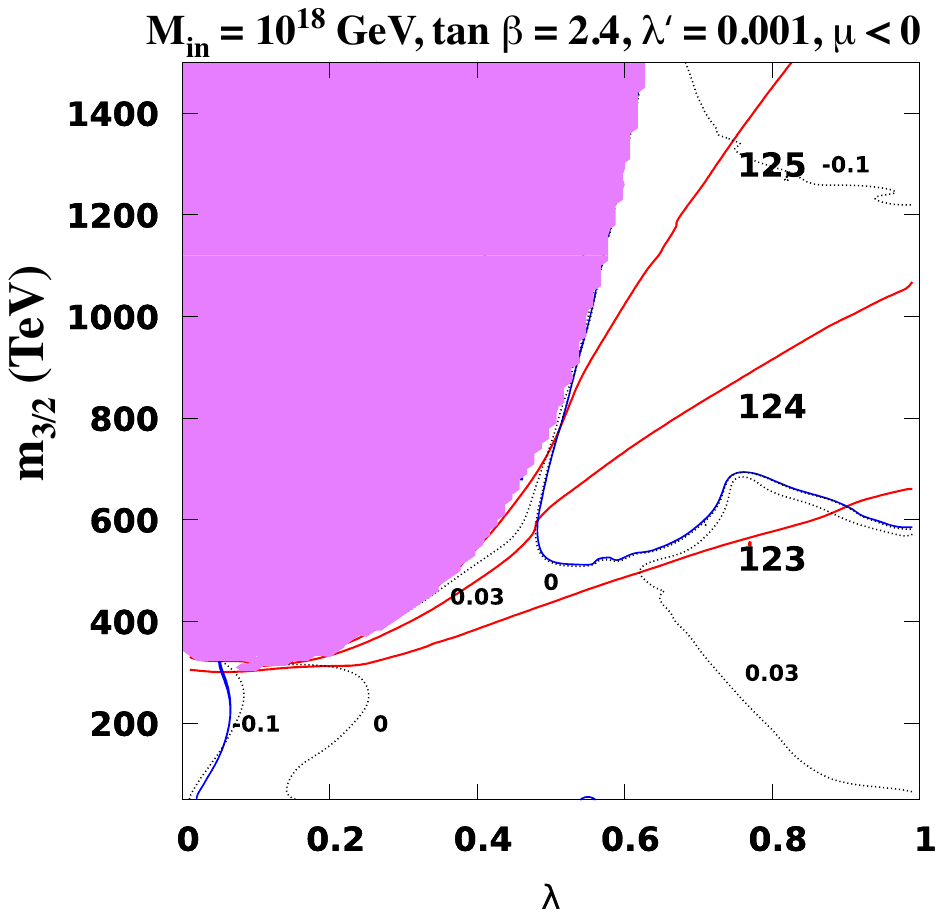}
  \vskip -1.2in
\caption{\it Examples of the $\lambda$-$m_{3/2}$ plane for $M_{\rm in} = 10^{18}$~GeV, and $\mu < 0$. In the upper panel, $\tan \beta = 3$ with $\lambda' = 1$, 
and in the lower panel,
$\tan \beta = 2.4$ with $\lambda' = 0.001$,
The solid red and dotted black curves show the Higgs mass $m_h$ and $c$, respectively. The solid green contours in the upper panel
show the value of the proton lifetime in units of $10^{35}$ years.
The blue shaded region corresponds to the areas where the LSP abundance agrees to the observed dark matter density. In the magenta region, we do not have successful rEWSB.
} 
  \label{fig:SuperPGM_1_24_18Con}
\end{figure}

A pair of ($\lambda,m_{3/2}$) planes for the case with $c \ne 0$
is shown in Fig.~\ref{fig:SuperPGM_1_24_18Con} with $M_{\rm in} = 10^{18}$~GeV
and $\mu < 0$. In the upper panel, $\tan \beta = 3$ and $\lambda' = 1$.
As in Fig. \ref{fig:SuperPGMnoe_1_24_18Con}, the magenta shaded
region corresponds to parameter values where rEWSM is not possible.
As we have seen before, this constraint forces one to relatively large
values of $\lambda$. For these parameter choices, $\lambda \le 1$
remains perturbative up to the input scale, $M_{\rm in}$.
The red contours correspond to the Higgs mass in GeV. The solid green contours correspond to the proton lifetime in units of $10^{35}$ years.
For the large value of $\lambda^\prime$ considered here,
the proton lifetime is close to but safely above the current experimental limit. Note that the fact that the proton lifetime 
is longer for smaller $m_{3/2}$ has to do with the fact that 
the gravitino mass controls the MSSM $\mu$ parameter (through EWSB) and hence the Higgsino masses, as well as the gaugino masses. When run up to the 
the GUT scale these affect the unification point and 
unified gauge coupling and hence the proton lifetime.
The black dotted curves show the values of $c$ calculated from the
gauge coupling matching conditions (\ref{eq:matchmhc}--\ref{eq:matchg5}). 
The blue shaded strip corresponds to the region where the 
relic density falls within 3$\sigma$ of the central value determined by Planck \cite{planck18}, which crosses the $m_h = 125$ GeV contour at 
$\lambda \approx 0.75$. We can compare the position of the endpoint of the relic density strip with $\lambda = 1$ with the result in Fig. \ref{fig:lampG3}a where we see that the 
horizontal line would be crossed by a contour with 
$m_{3/2}$ between 200 and 300 TeV in agreement with the endpoint in this panel.

In the lower panel of Fig. \ref{fig:SuperPGM_1_24_18Con} we
show the same plane with $\tan \beta = 2.4$ and $\lambda^\prime = 0.001$.
The Higgs masses are a bit lower, due to the decrease in $\tan \beta$.
We do not display the proton lifetime contours here because the lifetime 
is orders of magnitude larger than the experimental limit since 
$\lambda^\prime$ is small. Notice that there are two relic density strips in this 
case. In the strip with $\lambda \gtrsim 0.5$, the relic density 
strip corresponds to a 3 TeV wino mass as the sign of $c$ is changing (note the proximity of the relic density strip
to the $c = 0$ contour). This effect was seen in Fig. \ref{fig:lampG3} though for different values of $\tan \beta$.
When $\tan \beta < 3$ as it is here, the effect on the wino
mass is greater and occurs at a lower value of $\lambda^\prime$,
The relic density strip at low $\lambda \approx 0.07$ is due to
wino-bino coannihilation \cite{Baer:2005jq, Harigaya:2014dwa, Nagata:2015pra} where the LSP is bino-like for $\lambda < 0.07$ and wino-like at larger $\lambda$.
The bino (and wino) mass on the strip
with $m_{3/2} = 200$ TeV is approximately 2.1 TeV. 

\subsection{Models with K\"ahler-type non-renormalizable interactions}
\label{sec:all}

\begin{figure}[ht!]
  \centering
  \vskip -0.6in
 \includegraphics[width=0.58\columnwidth]{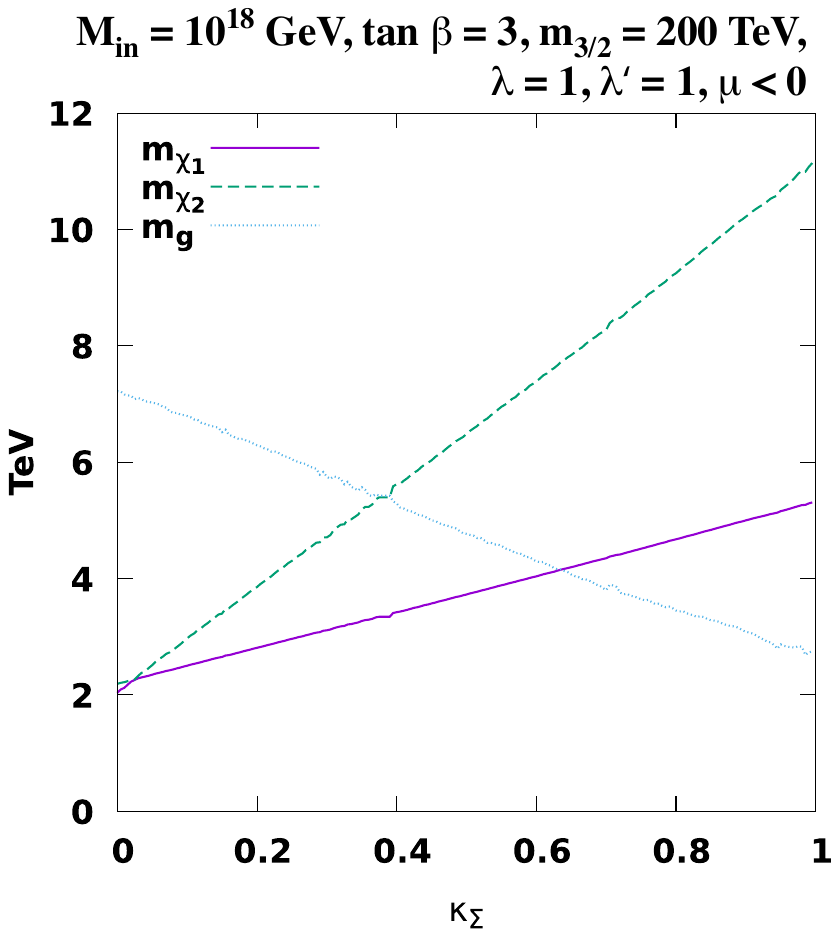}
         \hskip -1.2in 
 \includegraphics[width=0.58\columnwidth]{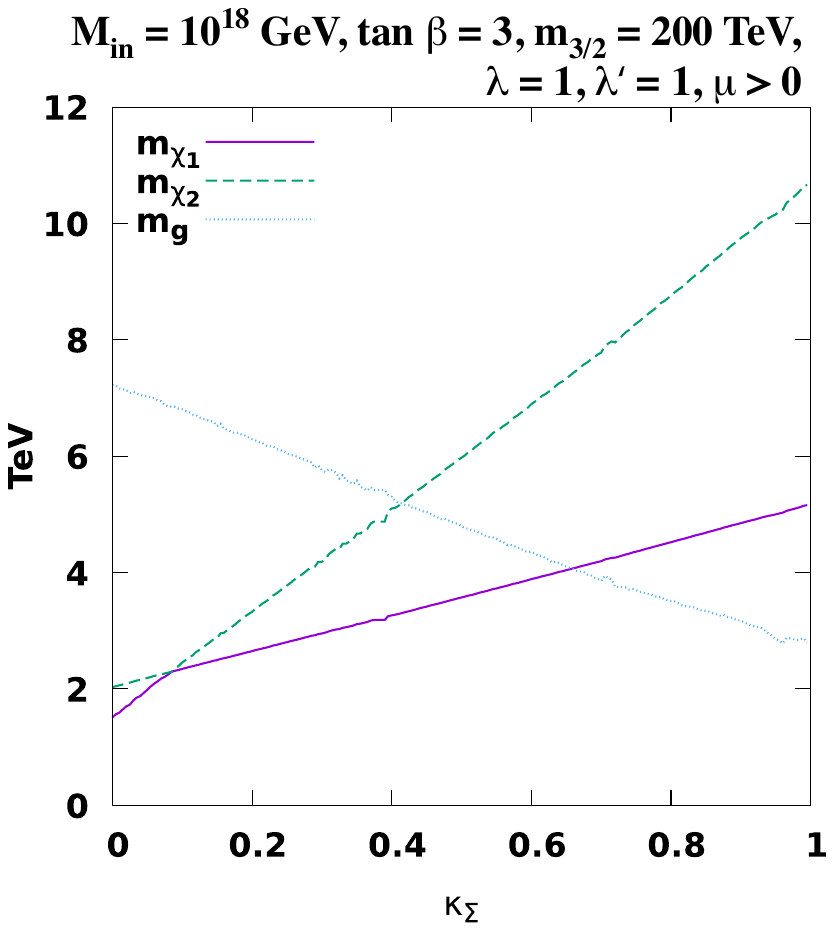}
 \\
\vskip -2in
 \includegraphics[width=0.58\columnwidth]{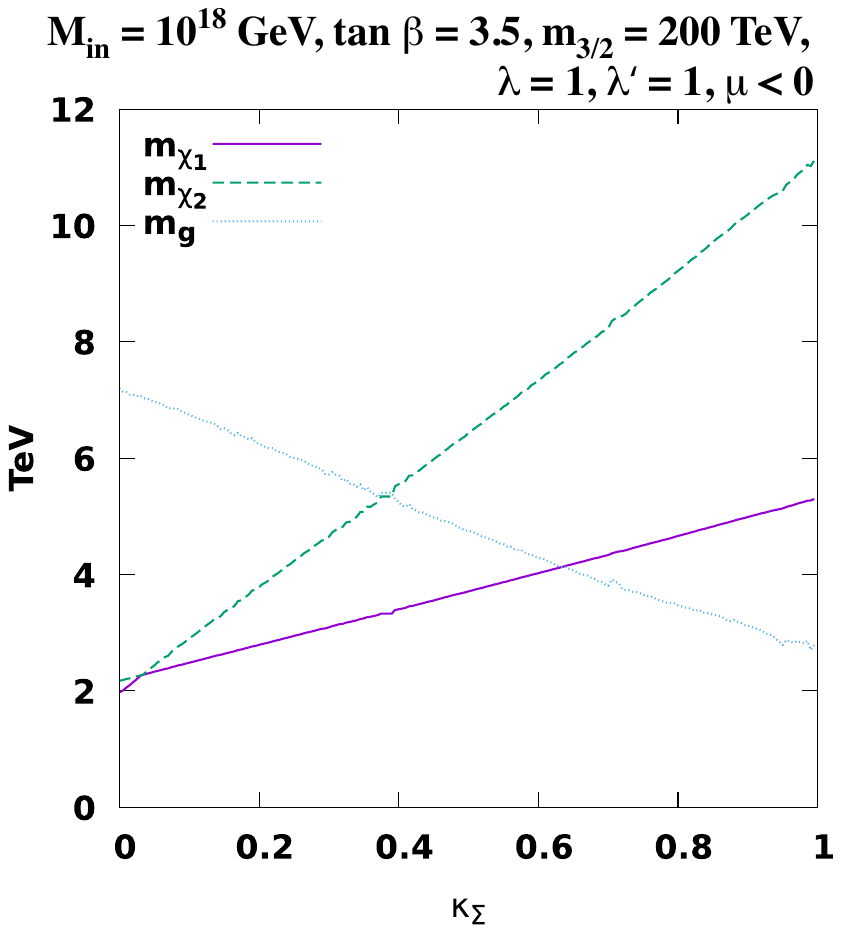}
         \hskip -1.2in 
 \includegraphics[width=0.58\columnwidth]{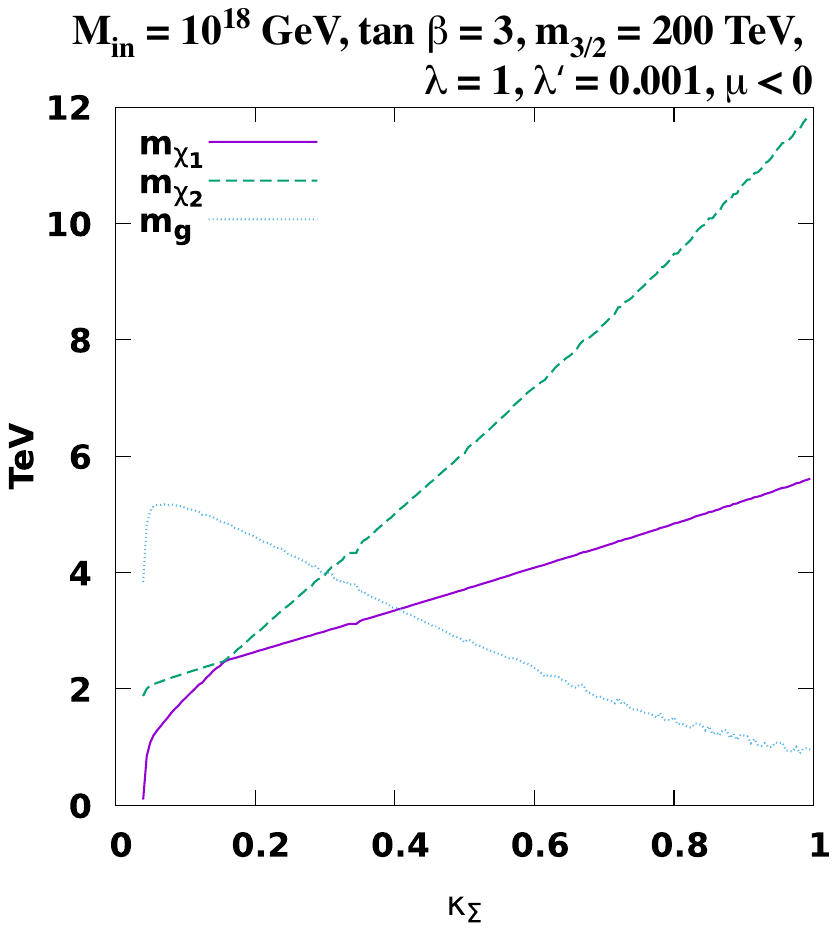}
 \vskip -1in
\caption{\it Gaugino masses as functions of $\kappa_\Sigma$ for
$M_{\rm in} = 10^{18}$~GeV, $m_{\rm 3/2} = 200 $~TeV and
 $\lambda = 1$.
In (a) $\mu < 0$, $\tan \beta = 3$ and $\lambda^\prime = 1$; (b) $\mu > 0$, $\tan \beta = 3$ and $\lambda^\prime = 1$;
(c) $\mu < 0$, $\tan \beta = 3.5$ and $\lambda^\prime = 1$;
(d) $\mu < 0$, $\tan \beta = 3$ and $\lambda^\prime = 0.001$.} 
  \label{fig:avsMi}
\end{figure}

Finally, we consider the effects of both the operators \eqref{eq:SigmaWW} and
\eqref{eq:delkz2}. In this case, the K\"{a}hler-type operators
\eqref{eq:delkz2} also modify the gaugino mass spectrum. To show the
significance of this effect, in Fig.~\ref{fig:avsMi},
we plot the gaugino masses as functions of $\kappa_\Sigma$ for
$M_{\rm in} = 10^{18}$~GeV,
$m_{\rm 3/2} = 200 $~TeV, and $\lambda = 1$. The
lightest (second lightest) neutralino mass is shown by the solid purple
(dashed green) line, while the blue dotted line shows the gluino
mass. In Fig.~\ref{fig:avsMi}a,
we have taken $\mu < 0$, $\tan \beta = 3$ and $\lambda^\prime = 1$.
For $\kappa_\Sigma = 0$, the wino is the LSP, which is replaced by the
bino for $\kappa_\Sigma \gtrsim 0.1$. For $\kappa_\Sigma \gtrsim 0.6$, the
gluino becomes the LSP and thus large values of $\kappa_\Sigma$ are 
phenomenologically disfavored. 
Near the cross-overs at $\kappa_\Sigma \simeq 0.1$ and
$\kappa_\Sigma \simeq 0.6$, the bino LSP is degenerate with the wino and
gluino in mass, respectively, and therefore we expect that coannihilation
\cite{Griest:1990kh} is quite effective. Indeed coannihilations with 
gluinos has been studied at length \cite{Profumo:2004wk, deSimone:2014pda, Harigaya:2014dwa, Evans:2014xpa,
Ellis:2015vaa, Nagata:2015hha} and allows 
LSP masses significantly higher than wino or bino masses in the absence of
coannihilation.

As seen in Fig. \ref{fig:avsMi}b, where we have
take $\mu > 0$, the sign of $\mu$ plays 
only a small role, affecting mainly the wino mass (at $\tan \beta < 3$
this effect becomes more pronounced). In Fig. \ref{fig:avsMi}c,
we consider $\tan \beta = 3.5$ with $\mu < 0$ and is very similar to the case with $\tan \beta = 3$.
In Fig. \ref{fig:avsMi}d, we have taken $\lambda' = 0.001$.
While the wino and bino masses are increased slightly, the gluino mass is decreased,
and becomes the LSP at lower $\kappa_\Sigma \approx 0.42$. We also see a sharp decrease
in the wino mass due to the effect of a sign change in $c$, already seen in 
Fig. \ref{fig:lampG2}.

\begin{figure}
  \centering
  \vskip -1in
  \includegraphics[width=0.58\columnwidth]{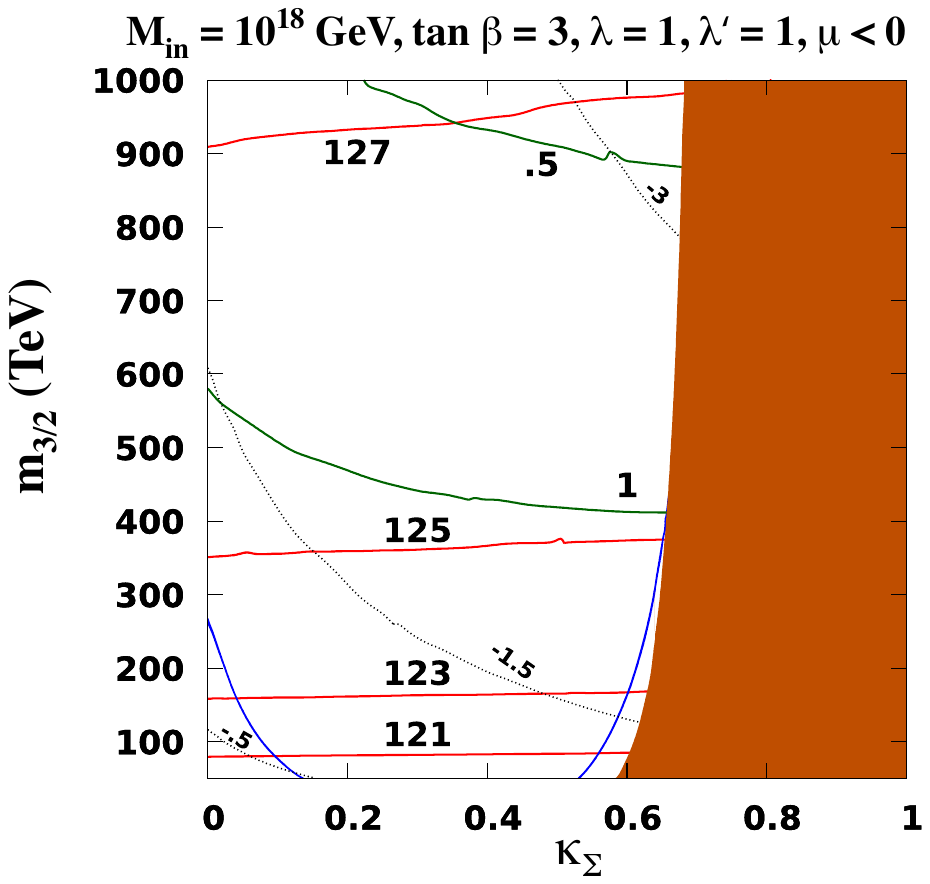}
        \hskip -1.1in 
  \includegraphics[width=0.58\columnwidth]{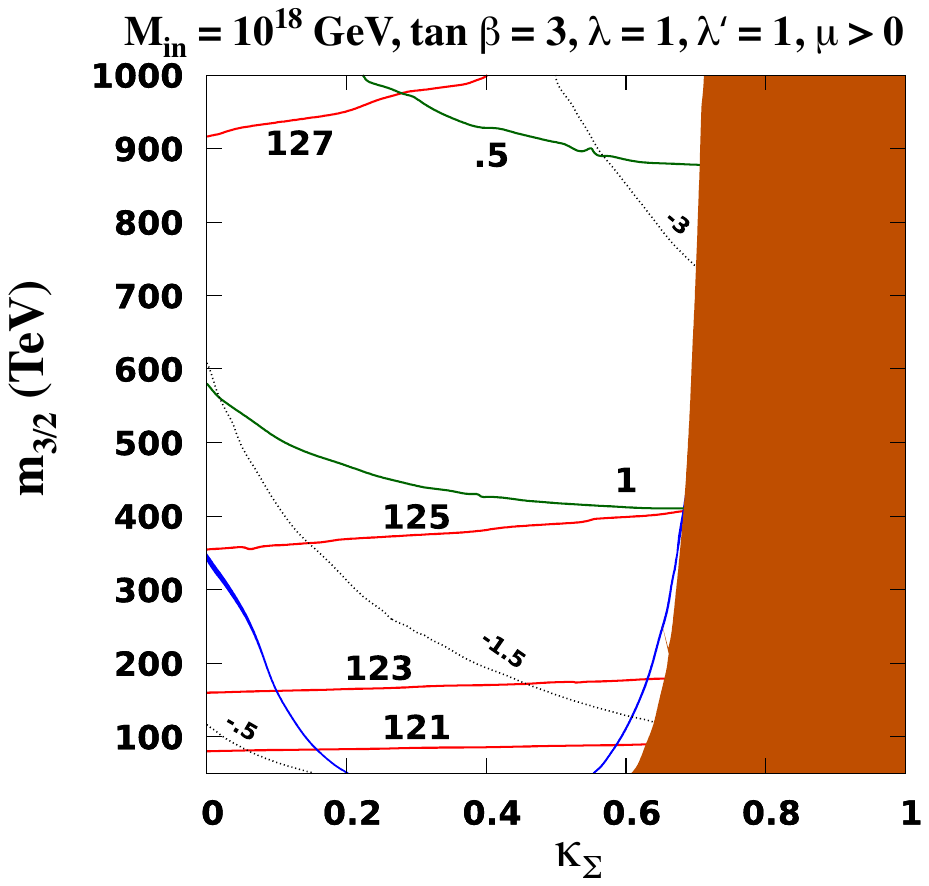}
  \\
\vskip -1.8in
  \includegraphics[width=0.58\columnwidth]{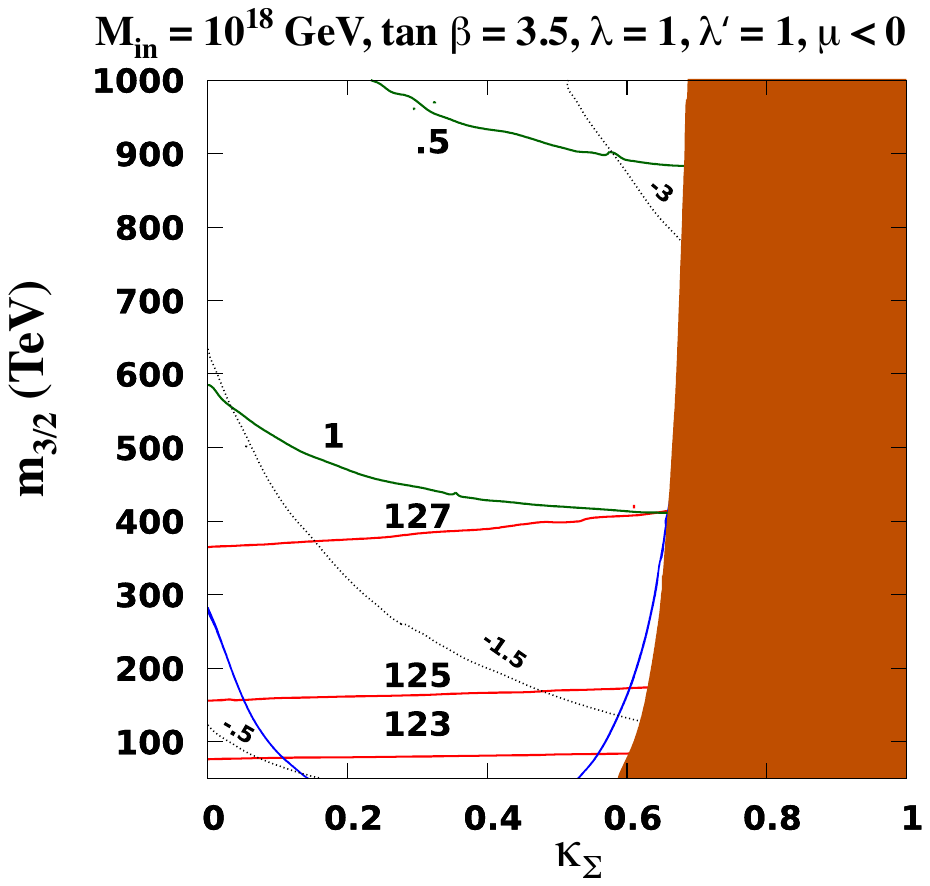}
        \hskip -1.1in 
  \includegraphics[width=0.58\columnwidth]{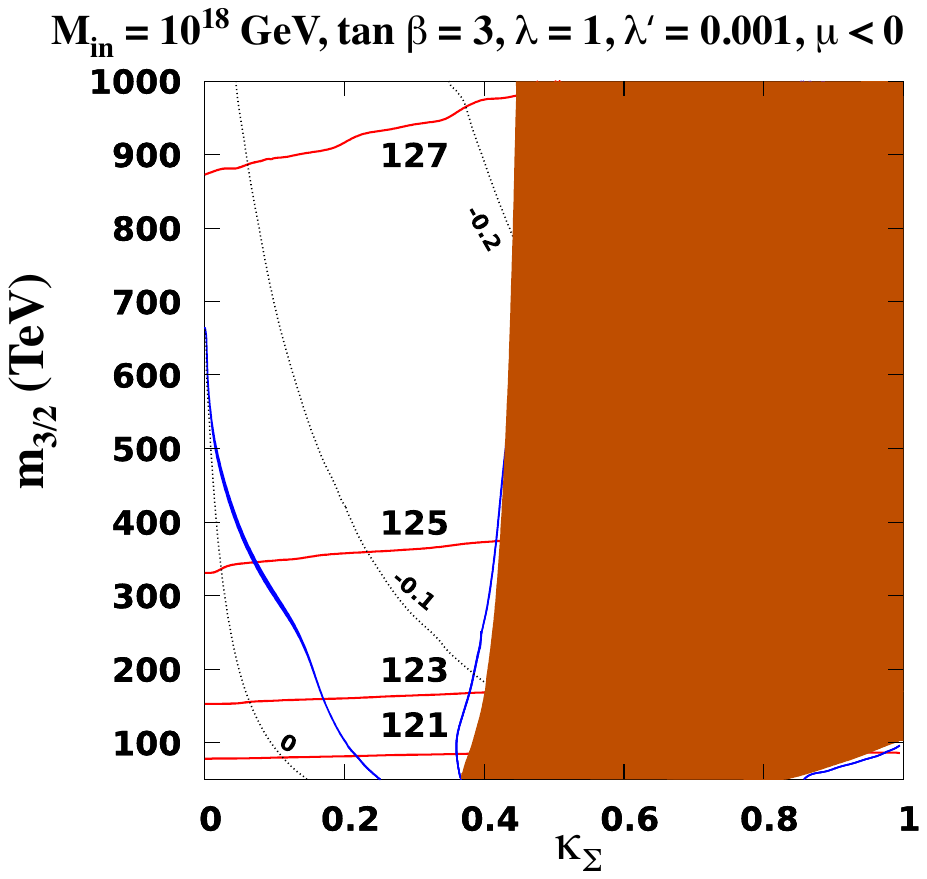}
  \vskip -1in
\caption{\it Examples of $\kappa_\Sigma$-$m_{3/2}$ planes for
$M_{\rm in} = 10^{18}$~GeV and
 $\lambda = 1$.
In (a) $\mu < 0$, $\tan \beta = 3$ and $\lambda^\prime = 1$; (b) $\mu > 0$, $\tan \beta = 3$ and $\lambda^\prime = 1$;
(c) $\mu < 0$, $\tan \beta = 3.5$ and $\lambda^\prime = 1$;
(d) $\mu < 0$, $\tan \beta = 3$ and $\lambda^\prime = 0.001$.
  The red and green solid contours
 correspond to the Higgs mass $m_h$ and the proton lifetime $\tau (p \to
 \pi^0 e^+)$ in units of GeV and $10^{35}$~years, respectively. The blue
 shaded region shows the areas where the LSP abundance agrees with the
 observed dark matter density. In the brown shaded region, the
 LSP is colored. The black dotted lines show the value of $c$.
}  
  \label{fig:SuperPGMpmx}
\end{figure}

In Fig.~\ref{fig:SuperPGMpmx}a, we show a
$\kappa_\Sigma$-$m_{3/2}$ plane for $\text{sign}(\mu) < 0$,
$\tan\beta = 3$, $M_{\rm in} = 10^{18}$~GeV, and
$\lambda = \lambda^\prime = 1$, where the Higgs mass $m_h$ and the
proton lifetime $\tau (p \to \pi^0 e^+)$ are plotted in units of GeV
and $10^{35}$~years in the red and green solid curves,
respectively. The black dotted curves indicate the values of $c$ (recall again that 
$c$ is calculated when $\lambda'$ is specified). The thin blue
shaded region shows the areas where the LSP abundance agrees to the
observed dark matter density to within 3$\sigma$. 
In the brown shaded region, the
LSP is gluino and thus is phenomenologically
disfavored. 
The gaugino masses for this case with $m_{3/2} = 200$ TeV were shown in
Fig. \ref{fig:avsMi}a.
We see that there are two regions where both the dark
matter density and the Higgs boson mass are explained. The region
with $\kappa_\Sigma \simeq 0$ is quite similar to that
discussed in Sec.~\ref{sec:wc}, and thus the LSP is wino-like.
On the other
hand, in the region where $\kappa_\Sigma \simeq 0.6$, the correct
relic abundance of the LSP is achieved via bino-gluino coannihilation
\cite{Profumo:2004wk, deSimone:2014pda, Harigaya:2014dwa, Evans:2014xpa,
Ellis:2015vaa, Nagata:2015hha}. From Fig. \ref{fig:avsMi}a, we see that bino and gluino
masses for this choice of parameters is about 4 TeV and may 
be just beyond the current reach of the LHC.\footnote{In this case, the gluino tends to be metastable because of the mass degeneracy between the bino and the gluino as well as some of the 
heavy sfermion masses. Such a gluino can efficiently be probed in displaced-vertex (DV) searches \cite{Nagata:2015hha}, while the sensitivity of the ordinary gluino searches based on the hard jets plus missing energy signals get worse due to the mass degeneracy. Currently, the most stringent bound is given by the ATLAS DV search based on the 32.8~fb$^{-1}$ data of the 13~TeV LHC: $m_{\tilde{g}} > 1550$--1820~GeV for $\Delta m = 100$~GeV ($\Delta m$ denotes the mass difference between gluino and the LSP) depending on the gluino lifetime \cite{Aaboud:2017iio}. This reach can be extended to $\sim 2$~TeV in the HL-LHC or up to 10 TeV in a 100~TeV collider. \cite{Ito:2018asa}.}
In these regions the proton lifetime is predicted to be $\gtrsim
10^{35}$~years; to probe this, therefore, we need at least 10-years (15-years) run at
Hyper-Kamiokande with a 372~kton (186~kton) detector \cite{Abe:2018uyc}.  

In Fig. \ref{fig:SuperPGMpmx}b,
we show the same plane with $\mu > 0$.
Comparing panels (a) and (b) of Fig.~\ref{fig:SuperPGMpmx},
we find that they are nearly identical except for the relic density strip
at low $\kappa_\Sigma$.  When $\kappa_\Sigma$ is small,
the LSP is a wino-like and more sensitive to the sign of $\mu$.
For $\mu > 0$ and a given gravitino mass, the wino mass is lower (relative to the case with $\mu < 0$), and we require a higher gravitino mass to obtain the same wino mass and hence relic density. This allows us to obtain the correct 
relic density with a Higgs mass closer to the experimental 
value of 125 GeV. In Fig. \ref{fig:SuperPGMpmx}c, we show the same plane with $\tan \beta = 3.5$. The only significant change is the 
value of $m_h$ which is roughly 2 GeV higher, making it easier to satisfy simultaneously the correct Higgs mass and relic 
density. The proton lifetime is hardly affected by the increase in $\tan \beta$.
Finally, in Fig. \ref{fig:SuperPGMpmx}d, we show the same 
plane as in panel (a) with $\lambda' = 0.001$. The lower value of $\lambda'$ causes a drop in the LSP mass (see e.g. Fig. \ref{fig:lampG2}) requiring an increase in the gravitino 
mass to obtain the same relic density.  As expected from Fig. \ref{fig:avsMi}d,
we see that the region excluded because the gluino is the LSP is larger extending
down to $\kappa_\Sigma \sim 0.4$.
In this case, the proton lifetime greatly exceeds the experimental lower limit
and no lifetime contours are shown. 
The other change apparent in panel (d) is the change in the value of $c$ and one even sees a contour with $c=0$ in the lower left corner of the figure. 
There is also a relic density contour which appears at large $\kappa_\Sigma$
and low $m_{3/2}$. As one can see from Fig.~\ref{fig:avsMi}d, as $\kappa_\Sigma$
increases the gluino mass decreases.  For low $m_{3/2}$ as in the right hand corner of this panel, the gluino mass passes through 0 and becomes large again
and for $\kappa_\Sigma > 0.8$ there is another possibility for 
bino-gluino co-annihilation (though $m_h \approx 121$ GeV here).

\begin{figure} [ht!]
  \centering 
  \vskip -1in
  \includegraphics[width=0.58\columnwidth]{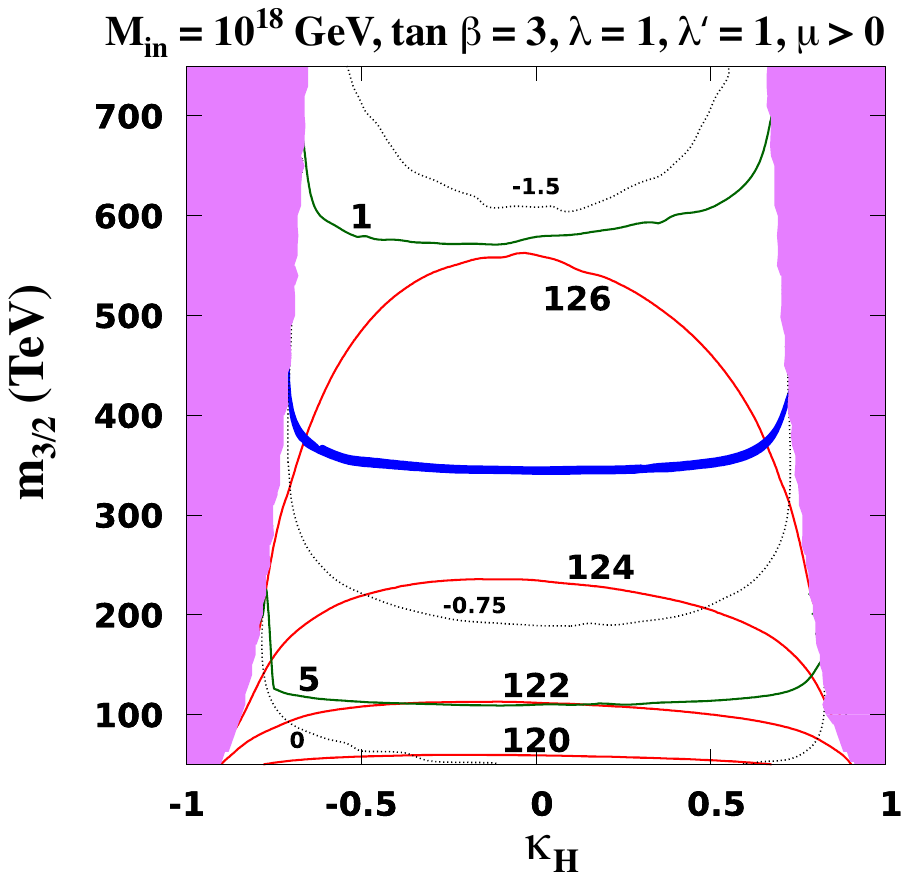}
  \vskip -1in
\caption{\it Example of a $\kappa_{H}$-$m_{3/2}$ 
plane for 
$\tan\beta = 3$, $M_{\rm in} = 10^{18}$~GeV,  
 $\lambda = \lambda^\prime = 1$, $\mu > 0$. The red and green solid contours
 correspond to the Higgs mass $m_h$ and the proton lifetime $\tau (p \to
 \pi^0 e^+)$ in units of GeV and $10^{35}$~years, respectively. The blue
 shaded region shows the areas where the LSP abundance agrees to the
 observed dark matter density. The black dotted lines show the value of $c$.
}  
  \label{fig:SuperPGMpmb}
\end{figure}

When $\kappa_H \ne 0$, 
the contribution to $m_H^2$ as given in 
Eq.~\eqref{eq:softinduced} makes it difficult to achieve
rEWSB. In Fig. \ref{fig:SuperPGMpmb}, we show an example of 
a  $\kappa_{H}$-$m_{3/2}$ plane for 
$\tan\beta = 3$, $M_{\rm in} = 10^{18}$~GeV,  
 $\lambda = \lambda^\prime = 1$, $\mu > 0$.
 Indeed when $|\kappa_H|$ is large, rEWSB is not attained as
 seen by the magenta shaded region. We see that for this choice of parameters,
 the relic density strip fall right about at $m_h = 125$ GeV. In this
 example, the proton lifetime is rather long, and is greater than $10^{35}$
 over much of the plane.

\begin{figure}[ht!]
  \centering 
  \vskip -1in
  \includegraphics[width=0.58\columnwidth]{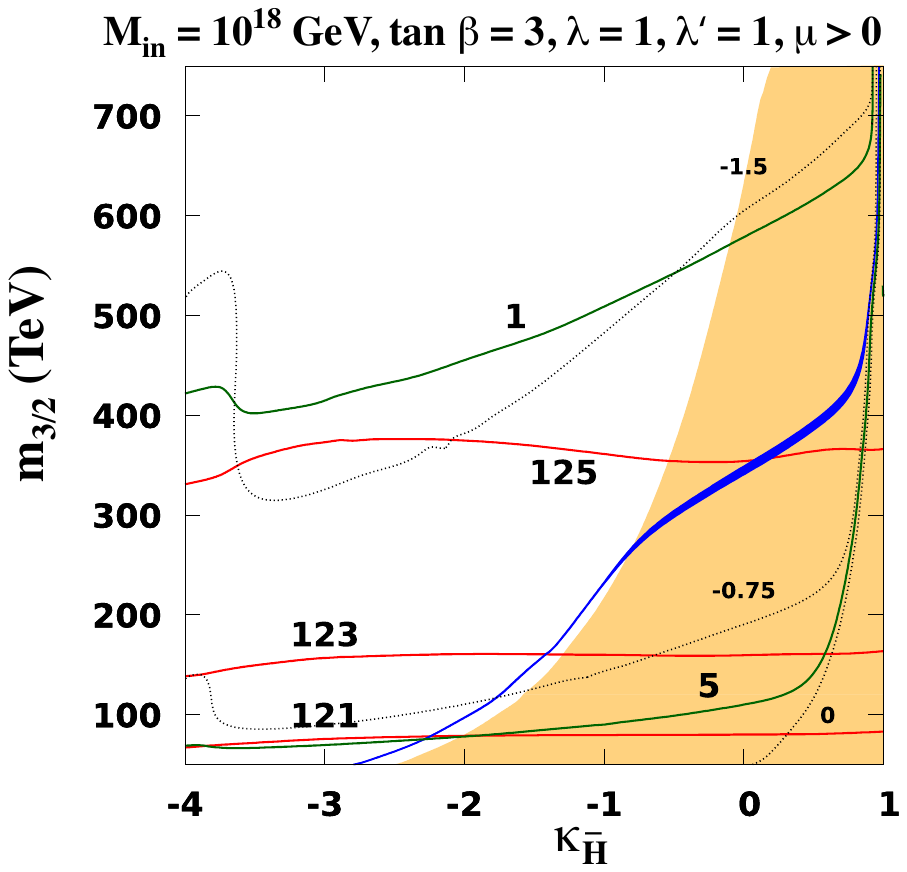}
       \hskip -1.1in 
  \includegraphics[width=0.58\columnwidth]{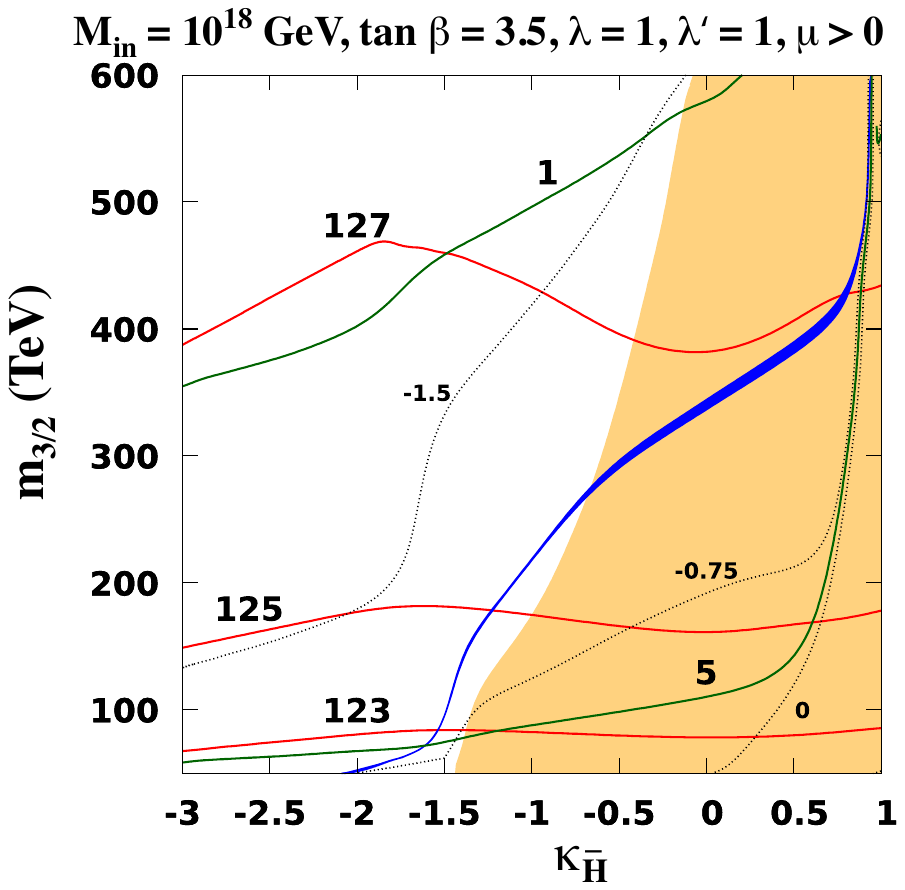}
  \vskip -1in
\caption{\it Examples of  $\kappa_{\bar{H}}$-$m_{3/2}$ 
planes for $\text{sign}(\mu) > 0$,
(a) $\tan\beta = 3$, and (b) $\tan \beta = 3.5$ with
$M_{\rm in} = 10^{18}$~GeV,  and
 $\lambda = \lambda^\prime = 1$. The red and green solid contours
 correspond to the Higgs mass $m_h$ and the proton lifetime $\tau (p \to
 \pi^0 e^+)$ in units of GeV and $10^{35}$~years, respectively. The blue
 shaded region shows the areas where the LSP abundance agrees to the
 observed dark matter density. The black dotted lines show the value of $c$. The orange regions is for a wino LSP and the white has a bino LSP.
}  
  \label{fig:SuperPGMpmy}
\end{figure}

Finally, in our last set of examples, 
we consider the effects of $\kappa_{\bar H}$.
In Fig.~\ref{fig:SuperPGMpmy}, we show a pair of
$\kappa_{\bar{H}}$-$m_{3/2}$ planes for $M_{\rm in} = 10^{18}$~GeV,  
 $\lambda = \lambda^\prime = 1$, $\mu > 0$. In the left panel, $\tan \beta = 3$
 and in the right panel $\tan \beta = 3.5$. In the orange shaded regions the 
 LSP is a wino, whereas elsewhere it is a bino.  Once again the proton lifetime is
 rather large and is greater than $10^{35}$
 over much of the plane. The Higgs mass increases by almost 2 GeV in the right 
 panel. The relic density strip at $\kappa_{\bar H} \lesssim -1$ is produced
 by wino-bino co-annihilations, whereas at larger $\kappa_{\bar H}$ it is simply a wino LSP with mass
 near 3 TeV.

\section{Conclusion and discussion}
\label{sec:conclusion}

While the LHC has nearly closed the door on ``low" energy
supersymmetry due to the absence of new (colored) scalars
or fermions with masses below $\sim 1$ TeV, we are left
with 2 nagging questions: Is supersymmetry a part of nature?
and if so, at what scale is it manifest? Of course,
the simplest way to definitively answer the first question,
is through discovery, and barring good fortune at future
runs of the LHC or its successor, both questions
will be difficult to answer. Nevertheless,
supersymmetry above the TeV scale may still have important
consequences on nature, such as providing a dark matter 
candidate and affecting proton decay. Like anomaly mediated 
supersymmetry breaking, pure gravity mediation contains massive scalars with gaugino masses of order 1 to several TeV.
In PGM, the scalars are invariably very heavy - of order
the gravitino mass which is typically 0.1 -- 1 PeV.

In constructing any top-down model, one must specify the scale
at which supersymmetry breaking is introduced. In the CMSSM
and similar models, this is usually chosen to be the GUT scale,
ie. the same scale at which gauge coupling unification occurs.
While there is no fundamental argument for this assumption,
it does remove one free parameter, making the model as constrained 
as possible. If the supersymmetry breaking input scale lies
above the GUT scale, the theory and low energy spectrum
becomes sensitive to GUT scale parameters, such as
the SU(5) Higgs self-couplings. In addition, 
higher order GUT operators may no longer be negligible and
may also affect the low energy theory.

In relaxing the assumption of GUT scale universality, one may, 
if the input scale is above the GUT scale, change the boundary conditions at the GUT scale. In effect, this allows us 
to start with mass universality at $M_{\rm in}$
and through renormalization group evolution, arrive at
non-universal masses at the GUT scale for the scalars.  
In the absence of higher order operators,
the gaugino mass spectrum is unaffected by running above the GUT scale as seen in Eq. (\ref{eq:amsbmssm}). Scalar non-universality
has an important effect on the Higgs soft masses in PGM which allows us to free up $\tan \beta$.
In anomaly mediated and PGM models, the gaugino
mass spectrum is directly related to the coefficients of the
one-loop beta functions. This leads to the lightest gaugino 
and hence LSP being a wino. As a consequence, to obtain
the correct relic density, the lightest gaugino must have 
a mass of $\sim 3$ TeV, which in turn fixes the gravitino mass.
In the superGUT version of PGM, the hierarchy of gaugino 
masses can be altered when the effects of 
higher order operators are considered. These higher dimensional operators generate A- and B-terms of order $m_{3/2}$. A- and B-terms this large give an order one correction to the matching conditions for the gauginos. This large deviation in gaugino matching conditions can lead to a bino or gluino LSP. At the boundaries between these regions with different LSP's, coannihilation is active and the bino can be a viable dark matter candidate. 

The higher dimensional operators also affect the gauge coupling matching conditions. As a result, the masses of the 
GUT scale gauge and Higgs bosons can be altered with
strong effects on the proton lifetime.

In this work, we have sampled a variety of superGUT PGM models
with and without the effects of higher order operators.
GUT scale PGM is highly constrained. It has two free parameters,
$m_{3/2}$ and $\tan \beta$ (when GM terms are included)
and while possible, the parameter space with
the correct relic density and Higgs mass is severely limited.
In the superGUT PGM, it is relatively easy to find a viable
parameter space so long as the Higgs coupling $\lambda$ 
is relatively large. This is largely due to the addition
of two new parameters, the supersymmetry breaking input scale,
$M_{\rm in}$ and the coupling $\lambda$.

At scales at or above the GUT scale, non-renormalizable,
Planck-suppressed operators may play a role in the low
energy spectrum.  Among these we considered first,
an operator coupling between the Higgs adjoint, $\Sigma$ and
the SU(5) gauge field strength with coupling, we denoted as $c$.
However, we used $c \ne 0$ to free up the Higgs adjoint
trilinear self coupling, $\lambda^\prime$ which was previously
constrained by GUT matching conditions. The theory now 
contains 5 parameters. The non-zero coupling, $c$
(or variable $\lambda^\prime$) alters the anomaly mediated
gaugino mass spectrum and makes it possible to obtain
a 3 TeV wino mass for a range in gravitino masses not possible with
$c=0$. It is also possible to find regions of parameter
space with a bino LSP with correct relic density
when bino-wino coannihilations are active.
Depending on the value of $c$ (or $\lambda^\prime)$,
the proton lifetime may be very large (as one might expect in PGM 
models) or potentially observable in on-going and future
proton decay experiments when $\lambda^\prime$ is large.

We also considered Planck-suppressed K\"ahler corrections for
the Higgs, {\bf 24}, ${\bf 5}$ and $\overline{\bf 5}$ representations with respective couplings, $\kappa_\Sigma$, $\kappa_H$, and $\kappa_{\overline{H}}$. These couplings
induce shifts in the $A$- and $B$-terms and hence affect
the gaugino matching conditions.  They also induce shifts in
the Higgs soft masses which affects radiative electroweak 
symmetry breaking. We have seen that the coupling
$\kappa_\Sigma$ has a strong effect on the gaugino mass.
Even a small non-zero value for $\kappa_\Sigma$ can cause 
$m_{\tilde B} < m_{\tilde W}$, and for larger values (of order
0.4 -- 0.6) can lead to a gluino LSP. When the wino/bino
or bino/gluino are nearly degenerate, coannihilations may
provide the correct dark matter density.

In absence of a discovery of physics beyond the Standard Model,
determining the scale of supersymmetry breaking will
indeed be challenging. As the mass scales associated with
supersymmetry increase, the likelihood of direct detection also
decreases. We may be reliant on more indirect signatures such 
as proton decay as discussed here, or other signatures 
if for example, $R$-parity is not exactly conserved 
and the LSP is allowed to decay with a long lifetime.


\section*{Acknowledgments}

The work of N.N. was supported in part by the Grant-in-Aid for Young
Scientists B (No.17K14270) and Innovative Areas (No.18H05542).
The work of K.A.O. was supported in part by the DOE grant DE-SC0011842 at the University of Minnesota.




\end{document}